\PassOptionsToPackage{table}{xcolor}
\documentclass[a4paper,11pt]{article}
\pdfoutput=1 
\RequirePackage{snapshot}  
\usepackage{jheppub} 
\usepackage{bm,amsmath,amssymb,slashed,graphicx,%
            enumerate,alltt,xspace,xcolor,multirow,mathrsfs}
\usepackage{fancyvrb}
\usepackage{booktabs,tabularx}
\usepackage{graphicx}
\usepackage{subcaption}
\usepackage{xspace}
\usepackage{cancel}
\usepackage[utf8]{inputenc}
\usepackage[compat=1.0.0]{tikz-feynman}
\usepackage{scalerel}
\usepackage{layouts}
\usepackage{adjustbox}
\usepackage{pdflscape}
\usepackage[normalem]{ulem} 
\usepackage{wasysym}
\usepackage{arydshln}
\usepackage{textcomp}
\usepackage{paralist}

\usepackage{algorithm}
\usepackage[noEnd=true]{algpseudocodex}

\newcommand{\ps}{\text{\textsc{ps}}}
\newcommand{\dl}{\text{\textsc{dl}}}

\newcommand{\nndl}{\text{\textsc{nndl}}}
\newcommand{\SL}{\text{\textsc{sl}}}
\newcommand{\nsl}{\text{\textsc{nsl}}}

\newcommand\betaps{\beta_\text{\textsc{ps}}}
\newcommand\betaobs{\beta_\text{obs}}

\newcommand{\mathd}{\mathrm{d}}

\newcommand{\drift}[1]{\ensuremath{\langle \Delta_{#1} \rangle}}
\newcommand\drifty {\drift{y}}
\newcommand\driftkt{\drift{\ln k_t}}

\newcommand\plhalf {PL$_{\beta=1/2}$}
\newcommand\pghalfd{PG$_{\beta=1/2}^\text{(disord)}$}
\newcommand\pgzerod{PG$_{\beta=0}^\text{(disord)}$}
\newcommand\pgzerosdfd{PG$_{\beta=0}^\text{(sdf,disord)}$}
\newcommand\pgzerosdf{PG$_{\beta=0}^\text{sdf}$}

\usepackage[capitalise]{cleveref}

\makeatletter
\g@addto@macro\bfseries{\boldmath}
\makeatother

\definecolor{labelkey}{rgb}{0,0.5,0.0}
\definecolor{royalpurple}{rgb}{0.47, 0.32, 0.66}

\usepackage{listings}
\definecolor{darkred}{rgb}{0.6,0.0,0}
\definecolor{darkgreen}{rgb}{0,0.4,0}

\definecolor{grey}{rgb}{0.5,0.5,0.5}
\definecolor{rust}{rgb}{0.9,0.4,0.0}
\definecolor{lightblue}{rgb}{0.0,0.5,1.0}
\definecolor{ForestGreen}{rgb}{0.13,0.55,0.13}

\allowdisplaybreaks 

\definecolor{semiblue}{rgb}{0.3,0.3,0.8}
\newcommand{\logbook}[2]{}

\newcommand{\pg}{{PanGlobal}\xspace}
\newcommand{\pl}{{PanLocal}\xspace}

\newcommand{\order}[1]{\mathcal{O}\left(#1\right)}
\newcommand{\as}{\alpha_s}
\newcommand{\aeff}{\alpha_{s,\text{eff}}}
\newcommand{\MSbar}{\ensuremath{\overline{\text{MS}}}}

\newcommand{\state}{{\cal S}}

\newcommand{\nc}{N_\text{\textsc{c}}} 

\newcommand{\calM}{\mathcal{M}}

\newcommand{\itilde}{{\widetilde \imath}}
\newcommand{\jtilde}{{\widetilde \jmath}}

\PassOptionsToPackage{hypertexnames=false}{jheppub}

\tikzstyle{block} = [rectangle, minimum width=1.0cm, minimum height=0.75cm, thin, draw=black]
\tikzstyle{blob} = [circle, minimum width=0.5cm, thin, draw=black]
\tikzset{blackarrow/.style={-stealth, semithick, draw=black}}
\tikzset{connection/.style={inner sep=0,outer sep=0}}

\newcolumntype{C}{>{\centering\arraybackslash}X}

\title{Higher logarithmic accuracy with disordered parton showers}
	
\preprint{{{\small MPP-2026-148, Nikhef 2026-013, OUTP-26-10P}}}

\newcommand{\OXaff}{Rudolf Peierls Centre for Theoretical Physics,
  Department of Physics, Parks Road,
  University of Oxford, Oxford OX1 3PU, UK}

\newcommand{\Torinoaff}{Dipartimento di Fisica, Università di Torino, and INFN, Sezione di Torino,
Via P. Giuria 1, I-10125 Torino, Italy}
\newcommand{\NIKHEFaff}{Nikhef, Theory Group, Science Park 105, 1098 XG, Amsterdam, The Netherlands}
\newcommand{\CNRSaff}{Université Paris-Saclay, CNRS UMR3681, CEA, Institut de physique théorique, 91191, Gif-sur-Yvette, France}

\newcommand{\UCLaff}{Department of Physics and Astronomy, University College London, London, WC1E 6BT, UK}
\newcommand{\MPPaff}{Max Planck Institute for Physics,
Boltzmannstr. 8, 85748 Garching, Germany}
\newcommand{\Monashaff}{School of Physics and Astronomy, Monash University, Wellington Rd, Clayton VIC-3800, Australia}

\author[a]{Melissa van Beekveld,}%
\author[b]{Silvia Ferrario Ravasio,}%
\author[c]{Keith Hamilton,}%
\author[d]{Alexander~Karlberg,}%
\author[e]{Gavin P.\ Salam,}%
\author[f]{Ludovic Scyboz,}%
\author[g]{Gregory Soyez}%

\affiliation[a]{\NIKHEFaff}
\affiliation[b]{\Torinoaff}
\affiliation[c]{\UCLaff}
\affiliation[d]{\MPPaff}
\affiliation[e]{\OXaff}
\affiliation[f]{\Monashaff}
\affiliation[g]{\CNRSaff}

\date{Received: date / Accepted: \today}

\abstract{ We show how the PanScales methodology for event-shape NNLL
  accuracy can be extended to a shower with dipole-local momentum
  recoil.
  A key issue is to obtain bounded splitting weights in the infrared, and a
  novelty that we introduce is to allow for disordering of emissions
  in a way that retains consistency with logarithmic accuracy.
  That same methodology can also bring significant speed improvements in the PanGlobal showers.
  Additionally, for the first time, we show how to consistently
  combine event-shape NNLL accuracy with the single-logarithmic
  treatment of spin correlations.
  The article shows the results of numerous leading-colour fixed-order
  and asymptotic tests.
  It also includes tabulated results for next-to-single (non-global)
  logarithms for energy flow in a slice with a finite number of
  flavours.
}

\keywords{QCD, Parton Shower, Resummation, LHC, LEP
  \\[4em]
  \textit{For the purpose of Open Access, the authors have applied a CC BY
  public copyright licence to any Author Accepted Manuscript (AAM)
  version arising from this submission.}
}

\begin{document}
\maketitle

\section{Introduction}

Parton showers are a central component of modern collider simulations.
By iterating elementary branching processes, they provide an exclusive
description of QCD radiation over a wide range of scales, from the
hard scale of the scattering process down to soft, non-perturbative
scales.  Their probabilistic and fully differential nature makes them
applicable to essentially arbitrary observables and event selections.
At the same time, these same features make the systematic improvement
of their perturbative accuracy considerably more challenging than for
inclusive calculations or analytic resummations.

A useful assessment of a parton shower's accuracy is
provided by its ability to reproduce logarithmically enhanced
contributions in regions containing multiple soft and collinear
emissions~\cite{Dasgupta:2018nvj,Dasgupta:2020fwr}.  In recent years,
this perspective has led to the construction of dipole showers that
achieve next-to-leading logarithmic (NLL) accuracy for broad classes of
observables, for example \cite{Dasgupta:2020fwr,Hamilton:2020rcu,
  vanBeekveld:2022zhl,vanBeekveld:2022ukn, vanBeekveld:2023chs,vanBeekveld:2026wnj} from our PanScales
collaboration, available as open source code~\cite{vanBeekveld:2023ivn}, and
similar developments by other groups, see for
example~\cite{Nagy:2020rmk,Forshaw:2020wrq,Herren:2022jej,Hoche:2024dee,Preuss:2024vyu,Helliwell:2025krf}.
A defining feature of these developments is that the
choice of evolution
variable, recoil prescription and phase-space mapping are treated as 
integral elements of the logarithmic structure of the shower, rather than
merely as implementation details.

Going beyond NLL accuracy requires the
inclusion of configurations in which two emissions are close to one
another in logarithmic phase space.
In the soft sector, the relevant configurations involve pairs of
partons with commensurate energies and angles.
Their matrix elements do not, in general, reduce to an iteration of
strongly ordered single-emission kernels.
The inclusion of such correlated matrix elements has been studied in
the past
\cite{Li:2016yez,Hoche:2017iem,Hoche:2017hno,Dulat:2018vuy,Gellersen:2021eci,vanBeekveld:2024qxs}.
In a programme of improving dipole shower logarithmic accuracy, particular
care is needed in how this inclusion is performed.
Such considerations were first discussed in
Ref.~\cite{FerrarioRavasio:2023kyg}, where double-soft contributions  were
incorporated at leading-colour accuracy%
\footnote{
In the following we will always work in the limit of a large number of colours, $N_c \to \infty$.
For the inclusion of subleading colour at NLL accuracy in the PanScales framework, see Ref.~\cite{Hamilton:2020rcu}.	
}
in a fully differential way into the PanScales showers.
The construction corrects the effective matrix element generated by
the shower to the exact leading-colour double-soft matrix element, including the
flavour and colour structure required to describe subsequent,
softer radiation.
This ingredient proved essential for achieving
next-to-single-logarithmic (NSL) accuracy for non-global observables~\cite{Dasgupta:2001sh} and next-to-next-to-double-logarithmic (NNDL) accuracy
for observables such as subjet multiplicities, accuracy that can be
validated against independent (semi-)analytic calculations in
Refs.~\cite{Banfi:2021owj,Banfi:2021xzn,Becher:2021urs,Becher:2023vrh,Becher:2026zon}
and \cite{Medves:2022ccw} respectively.

The same double-soft ingredient subsequently became part of the
framework developed in Ref.~\cite{vanBeekveld:2024wws}, which identified
the correspondence between the ingredients of analytic NNLL
resummations in the coherent branching formalism (see e.g.~\cite{Banfi:2004yd,Banfi:2018mcq}) and their parton-shower
counterparts.
Together with the required inclusive soft-collinear and hard-collinear
corrections, it enabled the first demonstration of NNLL accuracy for a
broad set of global event-shape observables.
These results represented an important step towards a systematically
NNLL-accurate general-purpose event generator.

A key element of the strategy adopted so far was that the
parton shower remained responsible for populating the full emission
phase space.
In particular, emissions were generated sequentially and in decreasing
order of the shower evolution variable.
This organisation 
allows the higher-order correction to be localised to the region in
which it is required.

The requirement that the corrected configuration first be generated
through an ordered sequence of shower branchings is, however, not
fundamental and can be restrictive.
Some regions of phase space are accessible only through suppressed
shower histories, requiring sizeable overestimates in the veto
algorithm.
The shower must consequently generate and reject additional trial
branchings, while the evaluation of the correction may require the
reconstruction and summation of several alternative histories.
These operations introduce a significant computational overhead,
especially when higher-order kernels are applied repeatedly throughout
the evolution.
Furthermore, it is well known that typical dipole showers employing
local momentum conservation do not fully populate the two-emission
phase space under strict
ordering~\cite{Lopez-Villarejo:2011pwr,Li:2016yez,Fischer:2017yja}.
One might worry that this concern could extend to the double soft
region, though for the PanGlobal showers we considered in
Ref.~\cite{FerrarioRavasio:2023kyg} that was not an issue.

In this work, we explore the option of allowing disordered emission.
Specifically, as well as relying on dedicated multi-emission kernels generated
through ordered shower histories, our construction allows emissions to
be inserted in an order that does not follow the shower evolution
variable.
Disordered shower emissions have been explored in the past in the form
of ``smooth ordering'' for matching to multi-parton hard matrix
elements~\cite{Giele:2011cb}.
The novelty of our work here is that the way our disordering is
formulated ensures that unordered insertions are suppressed except
when two emissions are close in logarithmic
(Lund~\cite{Andersson:1988gp,Dreyer:2018nbf}) phase 
space.
This guarantees that the NLL accuracy of the shower is maintained and
that modifications from the disordered configurations enter only at NNLL.
In the case of one of the showers we explore, the PanLocal shower, this
reorganisation is critical for reproducing the full double-soft matrix
element with bounded weights.
For the PanGlobal family of showers, it enables double-soft
matrix-element corrections to be implemented with a lower
overhead, especially for the $\betaps=1/2$ ordering choice, where there
is almost an order of magnitude gain in event generation speed.

A second aspect treated in this work concerns spin correlations between successive collinear branchings, which are required to
reproduce the azimuthal structure of multi-parton matrix elements, and for
NLL accuracy~\cite{Dasgupta:2020fwr,Karlberg:2021kwr}. They are
implemented in the PanScales showers~\cite{Karlberg:2021kwr,Hamilton:2021dyz,vanBeekveld:2022zhl}
by propagating spin-density and decay matrices along the
branching tree~\cite{Collins:1987cp,Knowles:1988hu}.
Combining this spin treatment with the double-soft correction of
Ref.~\cite{FerrarioRavasio:2023kyg} has remained an open problem.

When spin correlations are enabled, the shower already produces the correct
azimuthal modulations in the strongly angular-ordered limit via spin-density
matrices. The effective shower matrix element used in
the denominator of the double-soft corrections~\cite{FerrarioRavasio:2023kyg} was,
however, derived
assuming azimuthally uniform shower generation.
We resolve this incompatibility by using the collinear projection
of the splitting amplitude when generating the branching azimuth,
while retaining the full physical kinematics and original amplitudes
when constructing and evolving spin density matrices.
This facilitates 
the derivation of the corresponding spin-dependent
effective shower matrix elements for the denominator of the double-soft
correction.

The outline of the paper is as follows.
In Sec.~\ref{sec:overview} we provide an overview of the PanScales NLL showers for $e^+e^-$ collisions, and we describe the original algorithm for the inclusion of double-soft corrections.
In Sec.~\ref{sec:disordered} we describe how to combine disordered emissions alongside double-soft corrections.
In Sec.~\ref{sec:from-ds-to-nnll} we then illustrate how to obtain the other ingredients necessary for NNLL accuracy for global event shapes, following the recipe of Ref.~\cite{vanBeekveld:2024wws}.
In Sec.~\ref{sec:ds-and-spin} we describe our revised treatment of spin correlations and show how the double-soft algorithm can be straightforwardly adapted to incorporate them consistently.
We validate the logarithmic accuracy of the resulting PanScales showers
through a combination of fixed-order (Sec.~\ref{sec:fo-tests}) and
all-order tests (Sec.~\ref{sec:ao-tests}), and show timings in Sec.~\ref{sec:timings}.
In Sec.~\ref{sec:conclusion} we present our conclusions.
A forthcoming companion paper explores phenomenological
considerations~\cite{PanPheno}.

\section{Recap of double-soft corrections in the PanScales showers}
\label{sec:overview}
In this section, we start with a brief overview of the PanScales showers for lepton collisions, which come in two variants, \pg~\cite{Dasgupta:2020fwr,FerrarioRavasio:2023kyg} and \pl~\cite{Dasgupta:2020fwr}. 
We will only consider massless partons and leading colour effects. 
We then review our technique for implementing double-soft matrix element corrections.

\subsection{The \pg and \pl showers}
The \pg and \pl showers are both dipole showers that iterate $2 \to 3$ splittings
of colour dipoles.
We will use indices $\itilde\jtilde$ for the dipole that splits, and
$ik$, $kj$ for the two resulting dipoles, with $k$ the index of the
radiated particle.
The splittings are performed sequentially in terms of an ordering variable $v$. 
The \pg family of showers has a single kinematic map per dipole, and
distributes the transverse-momentum recoil globally across the event.
The \pl family involves two distinct kinematic maps per dipole,
assigning transverse recoil to one or other of the ends of the dipole.
The end that receives the transverse recoil is called the ``emitter''.
For both showers, the splitting probability at NLL is given by
\begin{align}
\label{eq:split-prob}
\frac{d \mathcal P_{\rm ps}}{d \ln v} = \sum_{\{\itilde, \jtilde\} \in {\rm dip}} \int d\bar{\eta} \frac{d\phi}{2\pi}\frac{\alpha_s^{\rm eff}(v, \bar{\eta})}{\pi} \, \left[f(\bar{\eta})a_k P_{\itilde \to ik}(a_k) + f(-\bar{\eta})b_k P_{\jtilde \to jk}(b_k) \right],
\end{align}
where $v \sim k_t\, e^{-\betaps |\eta|}$ is the ordering variable, $\betaps$ is
a shower-specific parameter that defines the scaling of $v$ with respect to
the transverse momentum $k_t$ and the rapidity $\eta$ of the emission in the dipole frame, and $\bar\eta$ and $\phi$ are two
auxiliary variables. In particular, $\bar{\eta}$ is a rapidity-like
variable constructed such that $\bar{\eta} = 0$ produces an emission
that bisects the dipole in the event centre-of-mass (CM)
frame, while $\phi$ is an
azimuthal angle.
For the PanGlobal shower, we also have a ``split-dipole-frame'' (sdf)
variant~\cite{FerrarioRavasio:2023kyg}, where $\bar \eta=0$ is defined
to be the bisector in the dipole centre-of-mass frame.

The leading-order (LO) splitting functions are denoted by $P_{\itilde
  \to ik}(a_k)$ and $P_{\jtilde \to jk}(b_k)$, with $a_k$ and $b_k$
representing the light-cone momentum fractions of the emitted particle
along the pre-branching $\tilde{p}_i$ and $\tilde{p}_j$ directions,
respectively. 
The function $f(\bar\eta)$ is a monotonic partitioning function that satisfies $f(\bar\eta)\to 1(0)$ when
$\bar\eta\to \infty(-\infty)$ and $f(\bar\eta)+f(-\bar\eta)=1$. 
In practice, our choice differs between PanGlobal and PanLocal. 
More details can be found, e.g., in Ref.~\cite{Dasgupta:2020fwr}.
Note that in the case of the PanLocal shower, the two terms in square
brackets in Eq.~(\ref{eq:split-prob}) correspond to the two distinct kinematic maps.
Finally, $\alpha_s^{\rm eff}(v,\bar\eta)$ denotes the effective strong coupling, which is a function of the $\MSbar$ coupling $\alpha_s^{\MSbar}(k_t)$ and the shower-specific anomalous dimensions.
At NLL, it reads
\begin{align}
\label{eq:aseff-nll}
\alpha^{\rm eff}_s(v, \bar{\eta}) = \alpha^{\MSbar}_s(k_t)\left[1 + \frac{ \alpha^{\MSbar}_s(k_t)}{2\pi}K_1 \right]\,,
\end{align}
where $\alpha^{\MSbar}_s(k_t)$ includes two-loop running effects.
The argument of $\alpha^{\MSbar}_s$ must reduce to the transverse
momentum of the emission $k_t$ in the soft-collinear
limit.
The constant $K_1$ is the two-loop cusp anomalous dimension 
\begin{equation}
  K_1 = C_A \left(\frac{67}{18}-\frac{\pi^2}{6}\right)-\frac{10}{9} n_f T_R.
  \label{eq:K1}
\end{equation}
This factor provides the correct inclusive NLO rate for soft-collinear
emission~\cite{Catani:1990rr}.
Together with the two-loop running of $\alpha_s^{\MSbar}$, it ensures
NLL accuracy for the shower's Sudakov exponent.
To achieve higher-order logarithmic accuracy, it is necessary to modify both the parton-shower matrix element (that is, the splitting function), to account, for example, for a pair of soft-collinear emissions that are close in phase space, and the effective coupling constant.
In Sections~\ref{sec:ds} and~\ref{sec:disordered}, we focus on the former, whereas the latter will be discussed in Sec.~\ref{sec:from-ds-to-nnll}.

\subsection{Original formulation of double-soft matrix element corrections}
\label{sec:ds}
At NNLL accuracy, the parton shower must correctly describe
configurations in which two soft-collinear particles are produced at
commensurate angles, while being well separated from all the other
particles.
The corresponding QCD matrix elements can be found, for example, in Ref.~\cite{Catani:1999ss}.
A strategy based on applying a
matrix-element correction to the effective double-soft matrix
element of the shower was originally implemented for the \pg{}
shower (see Ref.~\cite{FerrarioRavasio:2023kyg}).
The original construction was, however, limited in scope in two respects.
First, it was formulated for the case in which the shower spin-correlation
framework of Refs.~\cite{Karlberg:2021kwr,Hamilton:2021dyz} is disabled.
Second, Ref.~\cite{FerrarioRavasio:2023kyg} addressed only the \pg{} case.
Instead, for the \pl shower, the splitting weights obtained after
the double-soft corrections were found to be unbounded, because of
regions of phase space that the standard ordered shower reaches only
with arbitrarily small probability.
Before proposing solutions to these two shortcomings, we 
summarise the original modifications required to supplement our 2-jet
NLO matched showers~\cite{Hamilton:2023dwb} with double-soft
corrections. This is done while preserving their NLL accuracy for
global event shapes, further achieving NNDL accuracy for observables
such as subjet multiplicities.

Consider a dipole with hard legs $a$ and $b$ that emits a soft gluon
$\tilde{g}$, followed by the emission of a second soft gluon leading
to a configuration with two soft gluons $g_1$ and $g_2$, i.e.\ we have
a splitting sequence $ ab \to ab\tilde{g} \to abg_1g_2$.
The momenta of $g_1$ and $g_2$ can each  differ significantly from $\tilde{g}$ if
$g_1$ and $g_2$ have commensurate transverse momentum and rapidity.
Conversely, since $a$ and $b$ are taken to be hard particles, their momenta are not significantly affected by the soft-gluon emissions.
Without loss of generality, we assume that the shower has taken a path
where $g_2$ is the last radiated parton, that it is associated with a shower scale $v_2$, and that it is emitted from the $(a,\tilde g)$ dipole.%
\footnote{In
    the case where the second gluon, $g_2$, is emitted from a
    gluon-gluon dipole, $(\tilde g_k,\tilde g_l)$, the soft pair can
    be taken as either $g_kg_2$ or $g_2 g_l$. In practice
    we choose as the soft pair the one with the smaller invariant
    mass.
}
The first soft gluon, $\tilde{g}$, was emitted from the $(a,b)$ dipole at a scale $v_1$.
For a double-soft gluon pair, this corresponds to the colour flow
$ag_2g_1b$ (which, henceforth, we will write as $a21b$).
We also allow the soft pair to be a $\bar q_2q_1$ pair, and we use the
labels $1$ and $2$ collectively for the two particles in the pair.

In general, given the four particle momenta $p_{a,2,1,b}$, we will
consider two permutations of the momenta of the soft pair.\footnote{In
  more general matching~\cite{vanBeekveld:2026vbp} we have found it
  useful to consider a wider set of permutations, but in the
  double-soft case, the energy hierarchy between $a$ and $b$ on one
  hand and $1$ and $2$ on the other limits the maximum number of
  viable permutations to two.}
For a gluon pair, these correspond equivalently to two colour orderings;
for a $q\bar q$ pair they exchange which momentum is assigned to
the quark and antiquark.
The opposite ($a12b$) ordering would correspond, for example, to radiating
particle $2$ from the $(\tilde g,b)$ dipole.

\begin{figure}
  \centering

  \includegraphics[width=0.9\textwidth]{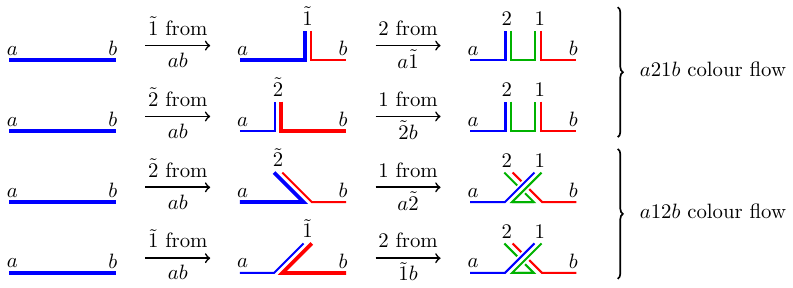}
  
  \caption{%
    \pg{} branching histories contributing to the emission of two soft
    gluons $(1,2)$ from a dipole with two much harder legs $(a,b)$.
    Lines represent colour-connected parton pairs, and the bold line
    indicates the dipole which is about to split. The first (last) two rows
    correspond to the case in which 2 is colour connected to $a$
    ($b$).
    In the first and fourth row, particle $1$ is taken to be the
    descendant of the first emission, while in the other rows particle
    $1$ is the second emission.}
  \label{fig:PG-histories}
\end{figure}

There are therefore four possible configurations, which we refer to as
states, $\state$, that can lead to the same set of momenta $p_{a,2,1,b}$:
$\state_{g_+}$, $\state_{g_-}$, $\state_{q_+}$, and $\state_{q_-}$.
The first index specifies the flavour of the emitted pair, either $gg$ or $q\bar q$, while the second specifies its ordering: `$+$' denotes the ordering initially produced by the shower, and `$-$' the inverted ordering.
For each state $\state_{i_\pm}$, one can compute the corresponding
exact-QCD double-soft matrix element, $\omega^{(E)}_{i_\pm}$,
and the effective shower matrix element, $\omega^{(S)}_{i_\pm}$.
The latter sums over all sequences of shower branchings
(``histories'') that lead to the same momenta, flavour assignment
and colour flow.
Without double-soft corrections, the shower therefore generates the following set of configurations:
\begin{equation}
	\state_S 	= \omega^{(S)}_{g_+} \state_{g_+} + \omega^{(S)}_{g_-} \state_{g_-} 	+ \omega^{(S)}_{q_+} \state_{q_+} + \omega^{(S)}_{q_-} \state_{q_-}.
\end{equation}
Our strategy is to apply a set of corrections that transforms this into the exact-QCD expectation:
\begin{equation}
	\state_E 	= \omega^{(E)}_{g_+} \state_{g_+} + \omega^{(E)}_{g_-} \state_{g_-} 	+ \omega^{(E)}_{q_+} \state_{q_+} + \omega^{(E)}_{q_-} \state_{q_-}.
\end{equation}
In a shower such as \pg{}, where the kinematic map has no distinction
between emitter and spectator, the radiation of two gluons $g_1$ and
$g_2$ with colour flow $a21b$ (state $\state_{g_+}$) could have been obtained through two
histories, which correspond to the first two rows of Fig.~\ref{fig:PG-histories}.
The first history is the one described above; in the second,
$g_1$ is emitted from the $(\tilde g,b)$ dipole and $g_2$ is
the recoiled $\tilde g$ gluon.
The last two rows of Fig.~\ref{fig:PG-histories} instead depict the
state $\state_{g_-}$, i.e.\ the opposite colour ordering. 
In a shower such as \pl{}, there is an additional choice of emitter
and spectator for each dipole, yielding a total of eight histories for
a given colour and flavour configuration.
Furthermore, for each history, $h$, the effective shower matrix element
must account for the requirement that the two emissions be correctly
ordered in the shower evolution variable $v^{(h)}$, i.e.\ a given
history with associated effective shower ordering variables $v_1^{(h)}$ and
$v_2^{(h)}$ should only be considered if $v_1^{(h)}\ge
v_2^{(h)}$.\footnote{When written with a superscript ${(h)}$ the index
  $i$ in $v_i^{(h)}$ refers to position in the sequence of emissions
  rather than a specific momentum index.}

With this at hand, we can proceed with describing our strategy to correct
for pairs of double-soft emissions in the shower.
For convenience, we introduce the exact $(E)$ and shower $(S)$ \emph{total} double-soft emission probabilities
\begin{subequations}
\label{eq:shower-weights}
\begin{align}
\omega^{(E)}
&= \omega^{(E)}_{g_+} + \omega^{(E)}_{g_-}
 + \omega^{(E)}_{q_+} + \omega^{(E)}_{q_-},
\\
\omega^{(S)}
&= \omega^{(S)}_{g_+} + \omega^{(S)}_{g_-}
 + \omega^{(S)}_{q_+} + \omega^{(S)}_{q_-}.
\end{align}
\end{subequations}
The explicit expressions can be found in
Appendix~\ref{app:shower-me}. A seemingly viable strategy would be to
separately reweight each of the terms in the sum
to the corresponding exact-QCD expectation as the associated configuration arises in the shower.
However, the ratios $\omega^{(E)}_{i_\pm}/\omega^{(S)}_{i_\pm}$ are not necessarily bounded for an NLL shower, even when particles 1 and 2 are widely separated in angle.
This is because an NLL shower is not required to produce the correct
yield of wide-angle $q\bar{q}$ pairs.  Similarly, the QCD matrix
element for a soft parton collinear to the hard parton $a$ but
colour-connected to parton $b$ is not always correctly reproduced by
the shower probability.  Instead, we first enforce the correct overall
probability, independently of the flavour and colour flow of the
states. Once a state is accepted, we either retain it as is, or change
its flavour or colour flow to reproduce the correct relative distribution
across the $\mathcal{S}_{i\pm}$ states.

In practice, we want to reweight the emission of
particle $2$, regardless of its flavour and colour connections, with
a factor $\omega^{(E)}/\omega^{(S)}$. We turn this into an emission
acceptance probability by introducing a constant overhead factor
$\Omega$, i.e.\ shower emissions are initially generated with a rate
multiplied by $\Omega$ and then accepted with a double-soft
acceptance probability,
\begin{align}
\label{eq:Pacc-ordered-only}
p_{\rm accept}
=
\frac{\omega^{(E)}}{\Omega\omega^{(S)}}.
\end{align}
The overhead factor $\Omega$ is chosen such that
$\omega^{(E)}<\Omega\omega^{(S)}$ is always satisfied.
Since the matrix elements associated with the shower histories are
evaluated in the double-soft limit, we expect
$p_{\rm accept}=1/\Omega$ (i.e.\ no double-soft correction,
$\omega^{(E)}=\omega^{(S)}$) whenever particles $1$ and $2$ are well
separated in the Lund plane (as long
as the shower includes spin correlations and these are treated coherently also
in the double-soft matching, Sec.~\ref{sec:ds-and-spin}).
This is an important requirement for preserving the NLL accuracy of
the shower, since the shower already gives the correct result in
such configurations.
It is only when the two emissions are within an $\order{1}$ distance
of each other in the Lund diagram that the double-soft matching brings
non-trivial corrections.

We now describe how to obtain the correct flavour and colour yields.
For this, we define the fractions
\begin{equation}
f_i^{(E,S)}
=
\frac{
\omega^{(E,S)}_{i_+}+\omega^{(E,S)}_{i_-}
}{
\omega^{(E,S)}
},
\qquad \text{and} \qquad
f_{\pm|i}^{(E,S)}
=
\frac{
\omega^{(E,S)}_{i_\pm}
}{
\omega^{(E,S)}_{i_+}+\omega^{(E,S)}_{i_-}
}\,,
\end{equation}
where $f_i^{(E,S)}$ represents the fraction of emissions with flavour $i$
and, for a given flavour $i$, $f_{\pm|i}^{(E,S)}$ is the fraction with a specified colour flow.
These quantities can be defined either for the uncorrected shower,
denoted by $(S)$, or for the exact-QCD expectation, denoted by $(E)$.

We next ensure that the flavour states are distributed correctly.
This is achieved by introducing \emph{flavour swaps}, i.e.\ by
relabelling $gg$ emissions as $q\bar q$ pairs, or vice versa.
In essence, when the parton-shower yield for a given flavour at fixed kinematics is lower than the exact yield, we do nothing. When it is larger, we change the flavour with a probability that depends on the excess. We therefore define the flavour-swap probability as
\begin{equation}
	p_{i \to j} = \max\left(1-\frac{f^{(E)}_{i}}{f^{(S)}_{i}},\,0\right),
	\label{eq:flav_swap}
\end{equation}
where $i$ is the current flavour and $j$ is the opposite flavour.
It is easy to verify that
\begin{equation}
  \label{eq:flavour-swap-sum-OK}
	(1-p_{i\to j})f^{(S)}_{i}+ p_{j\to i} f^{(S)}_{j} = f^{(E)}_i,
\end{equation}
i.e.\ we obtain the correct yield for flavour $i$.

The next step is to implement \emph{colour-flow swaps}
to guarantee the correct rates for the different colour flows.
If $i$ is the current flavour state, and $j$ is the flavour state
before a potential flavour-swap step (which may or may
not be $i$), we swap the colour connections with probability
\begin{equation}
  p_{\pm\to\mp|i(j)} = \max\left(1-\frac{f^{(E)}_{\pm|i}}{f^{(S)}_{\pm|j}},\,0\right).
\end{equation}
The swapping probabilities are such that, after accounting for events
that retain their original colour flow plus events that
are reassigned from the opposite flow, the fraction of
events in each colour flow matches the corresponding QCD prediction
(in analogy with Eq.~(\ref{eq:flavour-swap-sum-OK})).
This guarantees the correct QCD rate for every colour flow and
flavour.
Because these colour-flow swaps can be achieved by exchanging the momenta $p_1$ and $p_2$ of the soft pair, we refer to this step as either a colour-flow swap or a momentum swap.
Note that the colour-flow and flavour swap probabilities vanish  when
the two soft emissions are well separated in the Lund plane (again, as long
as the shower includes spin correlations and these are treated consistently also
in the double-soft matching).

\section{Disordered emissions}
\label{sec:disordered}

There are two critical conditions to check with the approach described
in Sec.~\ref{sec:ds}.
The first condition is that every phase space point (in the
double-soft region) is accessible through an allowed, ordered shower
history.
For the PanScales showers and $\betaps$ values that we use here, we
have verified that that condition is satisfied.

\begin{figure}
  \centering
  \includegraphics[width=0.49\textwidth]{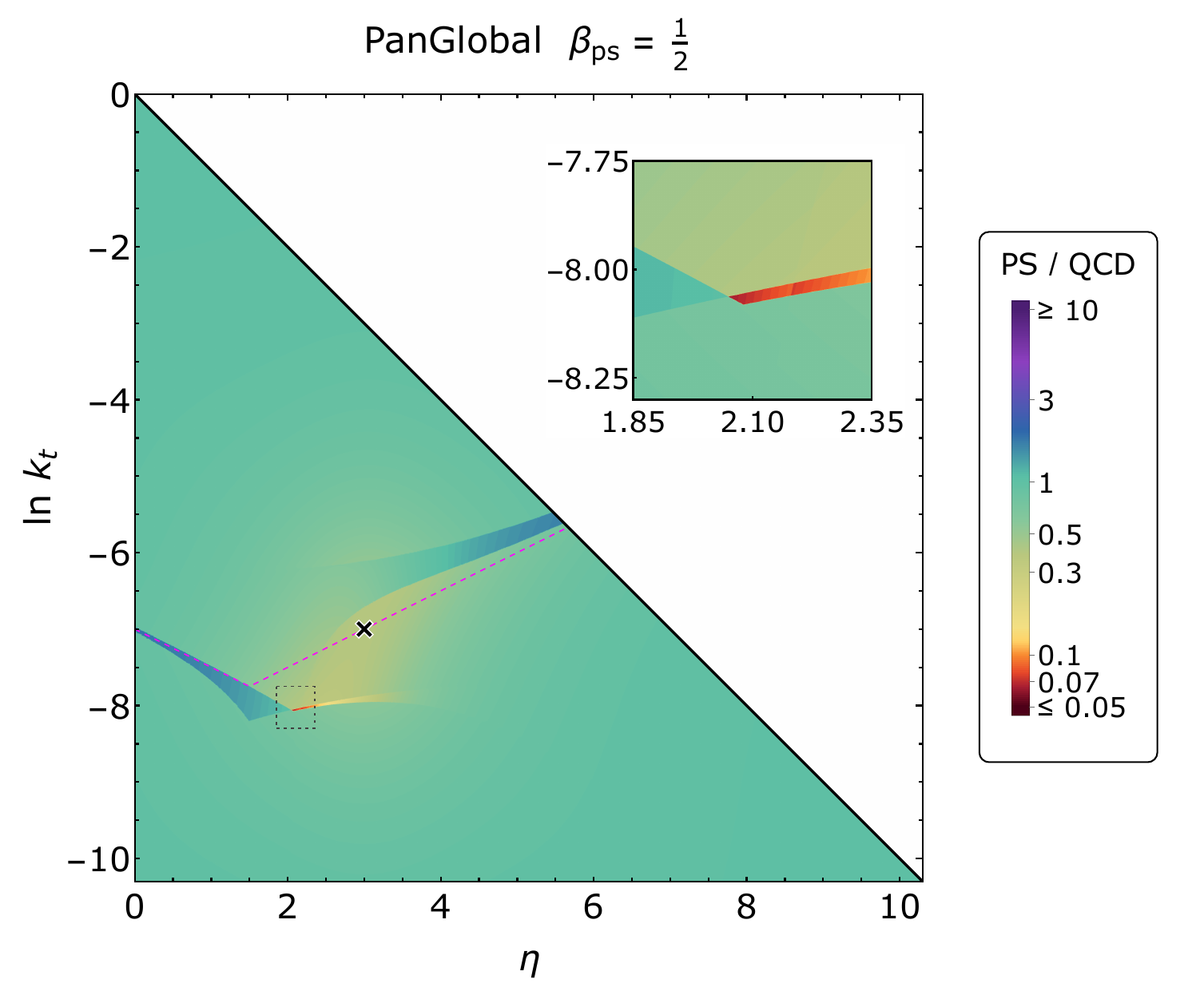}
  \includegraphics[width=0.49\textwidth]{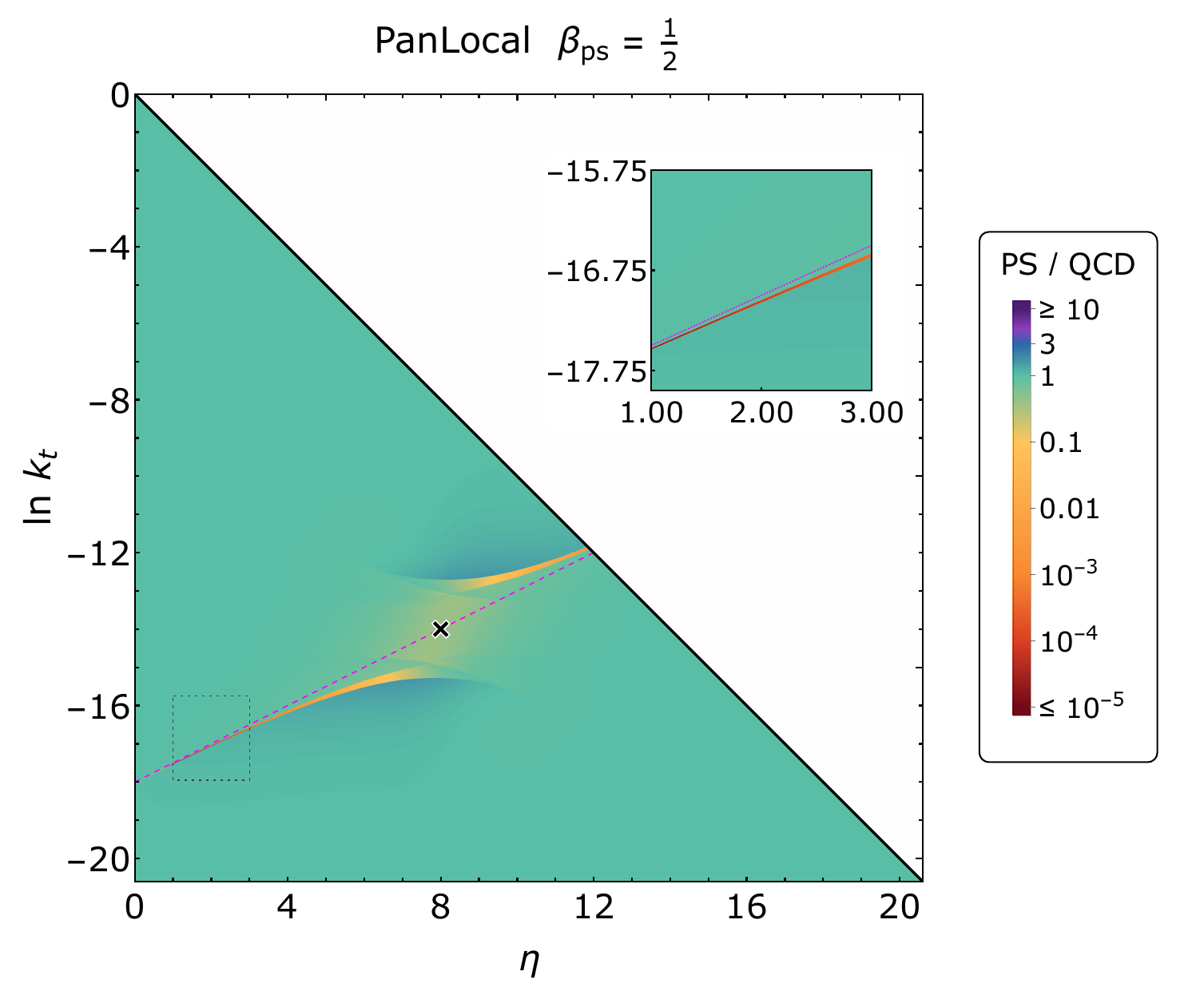}
  \caption{
    Ratio of the parton shower effective double-soft matrix element
    to the exact double-soft matrix element, for PanGlobal (left)
    and PanLocal (right), both with $\betaps=\frac12$.
    In each case the shower component is summed over all ordered histories,
    for $gg$ and $q\bar q$ channels ($n_f=5$).
    Coordinates are defined in the hard-dipole rest frame, with $p_a$
    in the $+z$ direction and transverse momenta in units where
    the mass of the $ab$ dipole equals one.
    For PanGlobal, the event centre-of-mass $4$-vector $Q^\mu$
    has been tuned to expose a
    situation where the ratio of shower to exact matrix element
    is especially small,
    ${Q^{\mu}}/{m_{Q}}\,=\,(\,49,\,0,\,50\sinh(3/2);50\cosh(3/2))/\sqrt{99}
    \simeq (\,4.9,\,0,\,10.7\,; 11.8)$.
    For PanLocal $Q^\mu\,=\,p_a+p_b\,=\,(\,0,\,0,\,0\,;1)$.
    The azimuthal separation of the fixed emission (the cross) and
    scanned emissions is 
    $\pi$, the azimuth of the former being aligned with $Q$ in the
    PanGlobal case.
    In the soft-collinear approximation, emissions on the magenta line
    that are far in rapidity from the fixed emission have the same
    evolution scale as that fixed emission.
  }
  \label{fig:PE-MS-ratio}
\end{figure}

The second condition is that there should be a finite overhead factor
$\Omega$ such that, for every phase-space point in the double-soft
region, the acceptance probability in Eq.~\eqref{eq:Pacc-ordered-only}
is always smaller than 1.
That condition is satisfied for the \pg shower, but not for the \pl
shower with strict ordering.
Fig.~\ref{fig:PE-MS-ratio} is intended to illustrate that.
It shows the ratio of the parton shower effective double-soft matrix
element to the exact QCD result, i.e.\ the inverse of
Eq.~\eqref{eq:Pacc-ordered-only} (without an overhead).
The configurations being studied involve one gluon that is fixed at
the rapidity and $\ln k_t$ indicated with the cross, while another gluon
is scanned over rapidity and $\ln k_t$.
No constraint is placed on which of the two gluons is produced first
in the shower, but the configuration must be reachable through an
ordered shower path.
Both colour orderings for the gluons are summed
over (and we also sum over soft $q\bar q$ production).
The gluons are chosen to have opposite azimuths, because that is the
most challenging configuration.

When the second gluon is far from the first, i.e.\ the bulk of the
Lund plane, the parton-shower over matrix element ratio in
Fig.~\ref{fig:PE-MS-ratio} is close to $1$, as expected.
However one sees non-trivial structures in regions of phase space in
the vicinity of the $\betaps=\frac12$ ordering line, and, in
particular, significant dips (yellow, red regions) to the left of the
cross, a little below the ordering line.
In the PanGlobal case the dip goes to a small finite value, which
translates to needing $\Omega \sim 20$.
In the PanLocal case we have found that the dip can go arbitrarily
close to zero, specifically as one considers more asymptotic kinematics
for the fixed gluon (note that the colour bar
stretches several orders of magnitude lower in range than for the \pg case).
That implies that there is no finite overhead $\Omega$ that will
guarantee bounded acceptance probabilities.\footnote{Specifically, the
  width of the ``fissure'' that is highlighted in the inset vanishes
  exponentially in the rapidity difference with the fixed emission,
  and inside the fissure the ratio of shower to exact matrix
  element also vanishes exponentially.
  This has been verified both numerically and analytically.
  \logbook{f3e35d21d3}{2024-05-29-disordered-emissions.tex, section 5
    discusses analytics. Look also into next versions of that section
    for additional discussions and plots.}
}

One of the main novelties of this paper is to show that one can circumvent this
limitation by allowing for disordered shower emissions, that is,
emissions that are no longer systematically ordered in decreasing
values of the shower ordering variable.
The algorithm described below has the important property that these
disordered emissions are allowed only in an order-one region of the logarithmic phase space, guaranteeing that we are not changing the behaviour of the shower when emissions are well-separated in the Lund diagram.
This ensures that the impact of disordering starts at NNLL order for event shapes.

\subsection{The algorithm}
\label{sec:algorithm}
We start by considering the emission of a pair of soft emissions, as in Sec.~\ref{sec:ds}.
However, instead of only allowing emissions that are strictly ordered in the evolution variable $\ln v$, we relax the strict shower ordering condition to
\begin{equation}
	\ln v_2 < \ln v_1+\Delta^\text{(disord)}\,,
\end{equation}
with $\Delta^\text{(disord)}$ an $\mathcal{O}(1)$ constant.
For a given configuration (phase-space point, flavour and colour
flow), one can compute the exact QCD expected rate as before (Eq.~\eqref{eq:shower-weights}).
However, in this case, we have two distinct shower contributions:
$\omega^{(S,\text{ord})}_{i_\pm}$ and $\omega^{(S,\text{disord})}_{i_\pm}$,
corresponding respectively to shower histories with $\ln v_2^{(h)}<\ln v_1^{(h)}$
and with $\ln v_1^{(h)} < \ln v_2^{(h)} < \ln v_1^{(h)}+\Delta^\text{(disord)}$ (the superscript~$^{(h)}$ is there to indicate that each history $h$ will have its own reconstructed
shower variables).
In practice, we thus have two types of configuration.
The first type involves standard {\it ordered} emissions, satisfying
$\ln v_2<\ln v_1$. 
These are treated as before, with the only
difference being that we saturate the corresponding global double-soft
acceptance probability as follows,
\begin{equation}\label{eq:disordered-Pacc-ordered}
	p_\text{accept}^\text{(ord)} = \min\left(\frac{\omega^{(E)}}{\Omega \omega^{(S,\text{ord})}},1\right),
\end{equation}
to guarantee that the acceptance does not exceed 1.
Note that here the shower effective weight only includes histories
with ordered configurations.
Besides this change, the flavour and colour-flow swaps remain
exactly as in Sec.~\ref{sec:ds}, keeping only the ordered histories;
$\omega^{(S)} \rightarrow \omega^{(S,\text{ord})}$ and
$\omega^{(S)}_{X} \rightarrow \omega^{(S,\text{ord})}_{X}$ for all
$X$.
In practice, this means that ordered emissions will produce
configurations of the form
\begin{align}
	\state_{S, \text{final}}^\text{(ord)}
	= \min\left(\omega^{(E)},\Omega \omega^{(S,\text{ord})}\right)\left[
	f^{(E)}_g\left( f^{(E)}_{+|g} \state_{g_+} + f^{(E)}_{-|g} \state_{g_-} \right)
	+ f^{(E)}_q\left( f^{(E)}_{+|q} \state_{q_+} + f^{(E)}_{-|q} \state_{q_-} \right)
	\right].
\end{align}
Instead, for {\it disordered} emissions ($\ln v_1^{(h)} < \ln v_2^{(h)} < \ln
v_1^{(h)}+\Delta^\text{(disord)}$), we apply the 
acceptance probability
\begin{equation}\label{eq:disordered-Pacc-disordered}
	p_\text{accept}^\text{(disord)}
	= \frac{\max(\omega^{(E)}-\Omega \omega^{(S,\text{ord})},0)}{\Omega \omega^{(S,\text{disord})}},
\end{equation}
where the contribution subtracted from the numerator only includes
{\it ordered} histories, while the denominator only includes {\em disordered} ones. 
Since Eq.~\eqref{eq:disordered-Pacc-disordered} is applied only when
$v_2^{(h)}>v_1^{(h)}$, the denominator always receives a contribution from at
least one history: the current one.
Flavour and colour-flow swaps are performed as in
the ordered case, except that they use the disordered flavour and
colour-flow shower rates.
This step produces a configuration
\begin{multline}
	\state_{S,\text{final}}^\text{(disord)}
	= \max\left(\omega^{(E)}-\Omega \omega^{(S,\text{ord})},0\right)
	\left[
	f^{(E)}_g\left( f^{(E)}_{+|g} \state_{g_+} + f^{(E)}_{-|g} \state_{g_-} \right)\right.\\
	\left.
	+ f^{(E)}_q\left( f^{(E)}_{+|q} \state_{q_+} + f^{(E)}_{-|q} \state_{q_-} \right)
	\right],
\end{multline}
such that the sum of the two types of emissions satisfies
\begin{align}
	\state_{S,\text{final}}^\text{(ord)}
	+ \state_{S,\text{final}}^\text{(disord)}
	= \state_{E}\,,
\end{align}
as required.
Through this procedure, we loosen the requirement of the ordered algorithm that $\omega^{(E)} \leq
\Omega \omega^{(S, {\rm ord})}$ to $\omega^{(E)} \leq \Omega
(\omega^{(S, {\rm ord})} + \omega^{(S, {\rm disord})})$ to be
satisfied for every point in phase space. 
Now $\Omega$ must be chosen such that Eq.~\eqref{eq:disordered-Pacc-disordered} is smaller than 1.

At this point, a few comments are in order. 
First, the procedure of capping the double-soft acceptance probability as in Eq.~\eqref{eq:disordered-Pacc-ordered} and adding a disordered contribution only affects a region of
order one in the Lund~\cite{Andersson:1988gp,Dreyer:2018nbf} diagram,
i.e.\ it has an effect which starts at the 
NNLL accuracy for event shapes. 
This is simply due to the fact that,
as long as the uncorrected shower is NLL-accurate,
$\omega^{(E)}/\omega^{(S,\text{ord})}$ (exponentially) tends to one when the two
emissions are separated in the primary Lund plane.

Secondly,  in practice, one could adopt different values for the overhead
factors for ordered and disordered emissions.
In this case, the ordered overhead factor should be used for the
ordered emissions as well as in the numerator of
Eq.~(\ref{eq:disordered-Pacc-disordered}), and the disordered overhead
factor should be used to generate disordered emissions as well as in
the denominator of Eq.~(\ref{eq:disordered-Pacc-disordered}).
For simplicity, in the results presented below, we have however chosen the same
overhead factor for both configurations.

Finally, even if, for a given parton shower, one can find an overhead
factor $\Omega$ such that Eq.~(\ref{eq:Pacc-ordered-only}) is
satisfied, it might be beneficial for speed reasons to adopt a smaller
overhead factor and allow for disordered emissions.
This is, for example, the case of the PanGlobal shower with
$\beta_{\textsc{ps}}=1/2$, for which ordered-only double-soft
corrections require an overhead factor $\Omega$ as large as 20.
Allowing for disordered emissions, one can reduce the overhead factor
to 2 (for $\Delta^\text{(disord)}=1$).
We discuss the impact of this reduction in Sec.~\ref{sec:timings} and
the associated Table~\ref{tab:timings}.

\subsection{Implementation in the Sudakov-veto algorithm}

To fully specify our strategy for using disordered emissions to correct
for double-soft emissions, we describe our practical implementation
and, in particular, how it is combined with the Sudakov-veto algorithm which is at the heart of the parton-shower multiple-emission strategy.
Starting from a given Born-level configuration, we set
$v_0=v_\text{start}$, the starting scale of the shower, typically,
$v_\text{start}=Q$ with $Q$ the centre-of-mass energy of the
$e^+e^-$ collision.
The double-soft algorithm is not applied to the very first emission, which is instead dealt with through matching corrections~\cite{Hamilton:2023dwb}.

We consider the generic situation where the parton shower has just generated
a normal (i.e.\ ordered) emission $k_i$ at a scale $v_i$.
The standard ordered shower would continue showering for scales
$v<v_i$.
The disordering algorithm instead proceeds as follows:
\begin{enumerate}
\item\label{algo-item-disord} If the last emission $k_i$ was a gluon, 
  use the usual Sudakov algorithm to potentially generate a
  disordered emission, starting from the shower scale
  $\min(v_i\exp(\Delta^\text{(disord)}),Q) > v_i$, with  $\Delta^\text{(disord)}$ of $\mathcal{O}(1)$.
  If an emission is attempted, accept it with a probability
  according to Eq.~\eqref{eq:disordered-Pacc-disordered}.
  This step is only performed on the two dipoles that contribute to
  the emitted $k_i$ gluon, requiring that the double-soft partner of
  the new candidate emission is $k_i$.\footnote{Failure to impose this
    last condition would lead to violations of the NNLL accuracy. This
    occurs in situations where we have a second gluon emission $k_2$
    at a shower scale $v_2$ with
    $\min(v_1\exp(-\Delta^\text{(disord)}),Q)<v_2<v_1$ and well
    separated from $k_1$ in rapidity.
    The situation of concern is when we explore the phase space region
    for an emission $k_3$ that is disordered with respect to 
    $k_2$.
    Specifically, it is important that we not double-count any of the
    double-soft phase 
    space region in the vicinity of $k_1$ that was already explored when attempting
    to generate $k_2$. \logbook{}{We have tested explicitly that this effect,
      while being small, was marginally visible in our NNLL accuracy
      tests.} }
  If the attempt was made at a scale $v > v_i$, the disordered
  procedure is repeated, with each step starting 
  where the previous one stopped, until either an emission is
  accepted, or one reaches the shower scale $v=v_i$ with no emissions
  accepted.
  In either case, go to step~\ref{algo-item-ord}.
  
\item\label{algo-item-ord}
  Run the standard Sudakov-veto
  algorithm down from the shower scale $v = v_i$.
  If an emission is attempted, accept it with a probability
  according to Eq.~(\ref{eq:disordered-Pacc-ordered}).
  If an emission is attempted but not accepted, simply repeat
  this step 2 from the new scale $v$ downwards.
  If instead the emission is accepted, we have generated a valid
  emission $k_{i+1}$ at a scale $v_{i+1}=v$ and one should then
  proceed to step~\ref{algo-item-disord}.
\end{enumerate}
One point to be aware of is that this algorithm introduces an
effective Sudakov factor into the double-soft real emission
generation (as does the original ordered algorithm).
Since the disordering covers an area in the Lund plane of order $1$,
the corresponding Sudakov factor has the structure $1- \order{\as}$,
which then multiplies the $\as^2$ double-soft matrix element.
The net effect of the Sudakov therefore appears only at order $\as^3$,
which is a N$^3$LL effect, i.e.\ beyond our accuracy.

\section{From double-soft corrections to NNLL accuracy for event shapes}\label{sec:from-ds-to-nnll}

General NNLL shower accuracy implies the following targets for
specific classes of observable 
\begin{compactitem}
\item NNDL accuracy ($\alpha_s^{n} L^{2n-2}$) for
  multiplicity-like observables;
\item NSL accuracy ($\alpha_s^{n} L^{n-1}$) for non-global observables primarily sensitive to soft emissions, such as the energy flow in a rapidity slice;
\item NNLL accuracy ($\alpha_s^{n} L^{n-1}$) for continuously global observables; 
\item NSL accuracy ($\alpha_s^{n} L^{n-1}$) for (single-logarithmic)
  collinear jet substructure observables like
  mMDT/SoftDrop ($\beta=0$, finite $z_\text{cut}$)~\cite{Dasgupta:2013ihk,Larkoski:2014wba}, fragmentation functions, and corresponding spin-sensitive observables.
\end{compactitem}
We will aim to achieve the first three of these at leading colour, leaving the last one to
future work. 

Schematically, to achieve higher logarithmic accuracy, we act on the
shower emission probability in two ways.
The work of Secs.~\ref{sec:overview} and \ref{sec:disordered} can be seen
as adding a factor $M(k)$ to Eq.~\eqref{eq:split-prob},
\begin{equation}\label{eq:emission-probability}
  \frac{\mathd \mathcal{P}_{\textsc{ps}}}{\mathd \ln v\,\mathd
    \bar\eta\, (\mathd \phi/2\pi)}
  = \frac{\aeff}{\pi} \times M(k)\times
  \left[
    a_kP_{\itilde \to ik}(a_k) f(\bar\eta) + b_kP_{\jtilde \to jk}(b_k) f(-\bar\eta)
  \right],
\end{equation}
with $M(k)$ engineered so as to yield the correct double-soft real matrix
element.
Additionally we need to upgrade the effective coupling,
$\aeff$, which becomes
\begin{equation}\label{eq:aseff}
  \aeff = \alpha_s^{\MSbar} \left[
    1 + \frac{\alpha_s^{\MSbar}}{2\pi}\left(K_1 + \Delta K_1(\bar\eta) 
      +B_2(z) + \Delta B_2(z) \right)
    + \left(\frac{\alpha_s^{\MSbar}}{2\pi}\right)^2\left(K_2 +
      \Delta K_2\right)
  \right],
\end{equation}
where $\alpha_s^{\MSbar}$ is evaluated at a scale corresponding to the
shower transverse momentum, with at least three-loop running.
The effective coupling includes a large-angle correction,
$\Delta K_1(\bar \eta)$, to the 2-loop cusp anomalous dimension, which
depends on the emission rapidity and azimuth, and the dipole opening
angle, as discussed in Ref.~\cite{FerrarioRavasio:2023kyg};
an NLO correction to the inclusive collinear $1\to2$ splitting function
$B_2(z)$~\cite{Dasgupta:2021hbh,vanBeekveld:2023lsa,Dasgupta:2022fim} (derived in Appendix~\ref{app:b2-nnll}),
supplemented by a shower-specific term $\Delta
B_2(z)$~\cite{vanBeekveld:2024wws,vanBeekveld:2024qxs}, 
which both depend on the momentum fraction $z$ of the emission;
and the three-loop cusp anomalous dimension $K_2$ in a
suitably defined scheme~\cite{Banfi:2018mcq,Catani:2019rvy}, together with a
shower-specific soft-collinear correction $\Delta
K_2$~\cite{vanBeekveld:2024wws}, which are both 
independent of the emission kinematics.
At our accuracy, the shower-specific contributions can be written in terms of
kinematic factors integrated over the 2-body double-soft phase space
(see Refs.~\cite{FerrarioRavasio:2023kyg,vanBeekveld:2024wws} for
further details and explicit expressions). 
Finally, we note that in the collinear region, the definition of the
transverse momentum to 
be used in the scale of the coupling is ambiguous (e.g.\ $zE\theta$ or
$(1-z)zE\theta$).  
The choice that one makes feeds into the expression for $B_2(z)$.

\begin{table}
\begin{center}
\begin{tabular}{lcccc|c}
\toprule
      observable & multiplicity & non-glob. & frag/substr. & ev.\ shapes  &  \\
      accuracy   & NNDL & NSL & \cellcolor{gray!15} NSL  & NNLL   & shower term \\
\midrule
      NLO matching           & needed & needed & needed & needed & $M(k)$ \\
      double-soft real       & needed & needed & needed & needed & $M(k)$ \\ \rowcolor{gray!15}    
      triple-coll. real      & - & - & needed & - & $M(k)$ \\
$\alpha_s^{(3\ell)}$         & - & - & - & needed & $\aeff$ \\
$B_2(z)$, $\Delta B_2(z)$    & - & - & \cellcolor{gray!15} needed & $\int \mathd z$ needed & $\aeff$ \\
$\Delta K_1(\bar \eta)$      & - & needed & - & $\int \mathd \bar\eta$ needed & $\aeff$ \\
      $K_2$, $\Delta K_2$    & - & - & - & needed  & $\aeff$ \\
\bottomrule
\end{tabular}
\end{center}
\caption{Contributions needed to achieve general NNLL shower accuracy,
  categorised according to the different classes of observable for
  which they are relevant.
  The last column indicates whether a contribution appears in the
  real branching probability, $M(k)$ in
  Eq.~(\ref{eq:emission-probability}), or in the effective coupling
  $\aeff$ of Eq.~(\ref{eq:aseff}).
  Elements in grey are not yet included in the PanScales showers.
  It is implicit that spin correlations should also be included,
  notably for collinear fragmentation and substructure observables.
  }\label{tab:contributions-needed}
\end{table}

Table~\ref{tab:contributions-needed} summarises the manner in which
each contribution affects the accuracy of various classes of
observable.
The elements in grey are those that we do not yet incorporate, and
that would be needed in order to achieve NSL accuracy for jet
substructure observables and fragmentation functions.

The shower-specific ingredients are distributed with the  PanScales
codebase.
The $\Delta K_1(\bar \eta)$, $\Delta B_2(z)$ and $\Delta K_2$ terms
all have their origin in the fact that shower branchings do not
conserve the same kinematic quantities as in analytic resummations.
In general, the quantity $\Delta K_1(\bar \eta)$ is in reality a
function also of the dipole opening angle and of the azimuth of the
emission with respect to the dipole plane.
It is provided as a 3D tabulation in the PanScales code.
The $\Delta K_2$ and integrated $\Delta B_2(z)$ contributions can be
determined from
the average ``drift'' in the logarithm of
transverse momentum ($\driftkt$) and in rapidity ($\drifty$) in the shower's $1\to 2$
branching of a soft-collinear emission, as compared to the conserved
quantity in the corresponding analytic calculation.
Specifically we have
\begin{equation}
  \int_0^1 \mathd z \frac{P_{q \to qg}(z)}{2C_F} \Delta B_2(z)
  = \frac{\beta_0\pi^2}{12}-\drifty -\driftkt,
\label{eq:dB2z}
\end{equation}
and
\begin{equation}
  \label{eq:DeltaK2-beta0}
  \Delta K_2 = -4\beta_0\driftkt,
  \qquad
  \beta_0 = \frac{11 C_A - 4 n_f T_R}{6}\,.
\end{equation}
The values of the drifts depend on the details of the kinematic map,
and in particular change relative to Ref.~\cite{vanBeekveld:2024wws} when using
disordering.
The values for the disordered showers are given in
Table~\ref{tab:drifts}.

\begin{table}
  \begin{center}
    \begin{tabular}{lclr@.lr@.lr@.l}
      \toprule
      shower & overhead ($\Omega$) & component & \multicolumn{2}{c}{$C_F$} & \multicolumn{2}{c}{$C_A$} & \multicolumn{2}{c}{$n_fT_R$} \\
      \toprule
      \multicolumn{9}{c}{spin correlations enabled} \\
      \toprule
      \multirow{2}{*}{\plhalf} & \multirow{2}{*}{3.0}
              & $\frac1{2\pi}\drifty$  &  0&79802(1) & 0&219439(2)  &  0&031818(1) \\
             && $\frac1{2\pi}\driftkt$ & -1&26217(1) & 0&751031(5)  & -0&059005(2) \\
      \midrule
      \multirow{2}{*}{PL$^{(\text{min})}_{\beta=1/2}$}& \multirow{2}{*}{3.0}
              & $\frac1{2\pi}\drifty$  &  1&13922(1) & 0&186594(2)  &  0&039214(1) \\
             && $\frac1{2\pi}\driftkt$ & -1&23300(1) & 0&765422(3)  & -0&059936(1) \\
      \midrule
      \multirow{2}{*}{\pghalfd}& \multirow{2}{*}{2.2}
              & $\frac1{2\pi}\drifty$  &  1&66743(1) & 0&016786(4)  &  0&052646(1) \\ 
             && $\frac1{2\pi}\driftkt$ & -1&74668(1) & 1&169692(7)  & -0&093461(2) \\ 
      \midrule
      \multirow{2}{*}{\pgzerod} & \multirow{2}{*}{2.2}
              & $\frac1{2\pi}\drifty$  &  0&05041(1) & 0&031934(4)  &  0&003836(1) \\
             && $\frac1{2\pi}\driftkt$ & -1&95981(1) & 1&176673(7)  & -0&104004(3) \\
      \midrule
      \multirow{2}{*}{\pgzerosdfd}& \multirow{2}{*}{2.2}
              & $\frac1{2\pi}\drifty$  & \multicolumn{2}{c}{0} & \multicolumn{2}{c}{0} & \multicolumn{2}{c}{0} \\
             && $\frac1{2\pi}\driftkt$ & -1&94625(1) & 1&164081(5)  & -0&103892(2) \\
      \toprule
      \multicolumn{9}{c}{spin correlations disabled} \\
      \toprule
      \multirow{2}{*}{\plhalf}& \multirow{2}{*}{3.0}
              & $\frac1{2\pi}\drifty$  &  0&79332(2) & 0&218003(2)  &  0&031929(1) \\
             && $\frac1{2\pi}\driftkt$ & -1&26366(1) & 0&749612(4)  & -0&059181(1) \\
      \midrule
      \multirow{2}{*}{PL$^{(\text{min})}_{\beta=1/2}$}& \multirow{2}{*}{3.0}
              & $\frac1{2\pi}\drifty$  &  1&13680(2) & 0&185055(2)  &  0&039398(1) \\
             && $\frac1{2\pi}\driftkt$ & -1&23457(1) & 0&764530(3)  & -0&060076(1) \\
      \midrule
      \multirow{2}{*}{\pghalfd}& \multirow{2}{*}{2.0}
              & $\frac1{2\pi}\drifty$  &  1&66608(3) & 0&018169(4)  &  0&052902(1) \\
             && $\frac1{2\pi}\driftkt$ & -1&75047(4) & 1&172262(11) & -0&094241(2) \\
      \midrule
      \multirow{2}{*}{\pgzerod}& \multirow{2}{*}{2.0}
              & $\frac1{2\pi}\drifty$  &  0&05291(1) & 0&030690(4)  &  0&004244(6) \\
             && $\frac1{2\pi}\driftkt$ & -1&96384(1) & 1&174562(6)  & -0&104581(2) \\
      \midrule
      \multirow{2}{*}{\pgzerosdfd}& \multirow{2}{*}{2.0}
              & $\frac1{2\pi}\drifty$  & \multicolumn{2}{c}{0} & \multicolumn{2}{c}{0} & \multicolumn{2}{c}{0} \\
             && $\frac1{2\pi}\driftkt$ & -1&95180(1) & 1&161880(4)  & -0&104683(1) \\
      \bottomrule
    \end{tabular}
  \end{center}
  \caption{Drift coefficients for the PanLocal shower with
    $\beta_{\textsc{ps}}=1/2$ and a double-soft overhead $\Omega=3$
    (allowing disordered emissions, as is always the case in this
    article for the PanLocal shower); for its
    PL$^{(\text{min})}_{\beta=1/2}$ variant using the ``minimal''
    scheme for the swaps introduced in
    Ref.~\cite{vanBeekveld:2026vbp}; and for three variants of the
    PanGlobal shower ($\beta_{\textsc{ps}}=1/2$,
    $\beta_{\textsc{ps}}=0$, and $\beta_{\textsc{ps}}=0$ in its
    ``split-dipole-frame'' (sdf) version), all three with a
    double-soft overhead $\Omega=2$ and allowing disordered emissions.
    In all cases, we work with a disordering buffer of
    $\Delta^\text{(disord)}=1$.
    The drift coefficients are split in colour channels and correspond
    to $2C_F=C_A=3$ and $n_f=5$.
    The numbers quoted include those colour factors.
    \logbook{ec5bf76648}{Fuller set of drift coefficients for a range
      of settings are given in logbook/summary-deltaK-drifts/}
  }\label{tab:drifts}
\end{table}

\section{Double-soft corrections and spin correlations}
\label{sec:ds-and-spin}

In a spin-averaged
shower, double-soft reweighting (Eq.~\eqref{eq:Pacc-ordered-only}) reproduces the physical azimuthal
correlation between the production plane of a soft intermediate gluon
and the splitting plane of that gluon, $g\to gg$ or $g\to q\bar q$.
Provided the parent gluon is soft, the correlation is physical; nevertheless,
in order to achieve the correct azimuthal correlation in \emph{all subsequent}
branchings, one must be able to include spin effects along the lines of the
spin-correlation tree algorithm of Ref.~\cite{Karlberg:2021kwr} in the double-soft-corrected shower.
The spin correlations modify the shower effective matrix element, $\omega^{(S)}$,
and to avoid double counting, the azimuthal structure generated by the shower must be taken into account in that factor.

As a reminder, the current implementation of spin correlations in the PanScales
code involves keeping track of helicity-dependent splitting amplitudes at each
stage of the branching, and storing them in a spin helicity tree.
For a branching $\tilde\imath\,\tilde\jmath \to i\,j\,k$, the branching amplitude used to
propagate spin information is given by~\cite{Karlberg:2021kwr,Hamilton:2021dyz}
\begin{align}
\label{eq:amp-old}
   A^{\lambda\,\lambda_1 \lambda_2} \, = \,
     - \sqrt{2} \, g_{s} \,
       \frac{\mathcal{F}^{\lambda\lambda_1\lambda_2}(z)}{ S_{-\tau}(i,k)}
       \left[
       \, 1 + \;
       \delta_{k,g} \, \delta_{\lambda,\lambda_1} 
       \left( \sqrt{\frac{1-z}{z}}\;
       \frac{S_{-\lambda_2}(i,j)}{S_{-\lambda_2}(k,j)} - 1 \right)
       \, \right],
\end{align}
where $\lambda$, $\lambda_1$ and $\lambda_2$ are the helicities of the
splitter before the branching, post-branching, and of the
radiation, $\tau$ is fixed by the parity of the summed helicities,
and $\mathcal{F}$ are helicity-dependent splitting amplitudes.
The $\delta_{k,g}$ factor indicates that the second term in square brackets
--- the soft-spin correction from Ref.~\cite{Hamilton:2021dyz} ---
is only present when the emitted parton is a gluon.
In practice, the spinor products $S$ are evaluated on the post-branching momenta and
built from an auxiliary function $\mathbb{S}$ which depends on a fixed reference
direction, i.e.~
\begin{subequations}
\begin{align}
\label{eq:S-def}
S_{\pm}(b,c) \, &= \, \sqrt{2 b\cdot c} \,\,\, \mathbb{S}_{\pm}^{*}(b,c) \, / \,|\mathbb{S}_{\pm}(b,c)|\,,\\
\label{eq:Sbb-def}
\mathbb{S}_{+}(b,c) \, &= \, \big(b^{x} + i\,b^{y}\big)\,(c\!\cdot\! k_0)
- \big(c^{x} + i\,c^{y}\big)\,(b\!\cdot\! k_0)\,,
\qquad
\mathbb{S}_{-} \, = \, -\,\mathbb{S}_{+}^{*}\,,
\end{align}
\end{subequations}
with $k_0$ the lightlike vector defining the spinor basis, equal to
$(1,0,0,-1)$ for a reference direction along $\hat z$.
 
Spin correlations occur as a result of a gluon emission with momentum fraction
$z_1$ that subsequently splits into a $gg$ or $q\bar{q}$ pair (with momentum fraction $z_2$). 
Once a set of generation variables $\ln v, \bar{\eta}$ is accepted, the shower
chooses the azimuth of the branching, $\phi$, based on the probability
\begin{align}
\label{eq:spin-accept}
p_{\rm spin} = \frac{|\calM(\phi | \{\tilde{p}\}, v, \bar{\eta})|^2}{|\calM_{\rm max}(\{\tilde{p}\}, v, \bar{\eta})|^2}
= \frac{\rho^{\lambda\lambda'}\,
A^{\lambda\,\lambda_1\lambda_2}(\phi)\,
A^{*\,\lambda'\lambda_1\lambda_2}(\phi)\,}{|\calM_{\rm max}(\{\tilde{p}\}, v, \bar{\eta})|^2}\,,
\end{align}
where $\rho^{\lambda\lambda'}$ is the spin-density matrix of the splitter,
$|\calM|^2$ is the amplitude squared obtained from evaluating the full spin
helicity tree, and $|\calM_{\rm max}|^2$ is a $\phi$-independent overestimate.
The expected physical form for $|\calM|^2$, whenever the current emission is collinear, is given by
\begin{align}
\label{eq:expected-ABC-modulation}
|\calM(\phi | \{\tilde{p}\}, v, \bar{\eta})|^2 \propto \left[ 1 + \mathcal{A}(z_1)\mathcal{B}(z_2)\cos(2\phi)\right] \,,
\end{align}
with the functions $\mathcal{A}$ and $\mathcal{B}$ given in
Refs.~\cite{Karlberg:2021kwr,vanBeekveld:2022zhl}.
For a double-soft configuration (i.e.~a soft intermediate gluon, $z_1 \to 1$,
$\mathcal{A}(z_1) \to 1$),\footnote{
Importantly, the presence of a soft particle leads to the factorisation of any
soft spin correlation tree --- because the emission of soft gluons is independent
of the helicities of the hard emitting partons.
This implies that one can indeed consider pair-wise histories for the
computation of the shower's double-soft effective matrix element.
}
we expect the effective matrix element with spin
corrections enabled to be of the form
\begin{align}
\label{eq:expected-mod}
\omega^{(S)}_{\mathrm{spin}} = \omega^{(S)}\left(1 + \mathcal{B}_{kg}(z_2) \cos(2 \Delta \psi_{12})\right),
\end{align}
with $\mathcal{B}_{kg}(z_2)$ encoding the modulation for $g\to gg$ or
$g \to q \bar q$ splittings, $k\in \lbrace q, g \rbrace$, and $\Delta \psi_{12}$
the measured azimuthal angle between the two successive splitting planes.
The functions $\mathcal{B}_{kg}(z_2)$ are given by~\cite{Karlberg:2021kwr}
\begin{align}
	\label{eq:Bkg}
	\mathcal{B}_{gg}(z_2) &= \frac{z_2^2(1-z_2)^2}{(1-z_2(1-z_2))^2}\,, \quad
	\mathcal{B}_{qg}(z_2) = \frac{-2z_2(1-z_2)}{1-2z_2(1-z_2)}\,.
\end{align}
Refs.~\cite{Karlberg:2021kwr,Hamilton:2021dyz} show that with spin correlations
enabled, the PanScales showers precisely reproduce
the pure cosine azimuthal modulation of Eqs.~\eqref{eq:expected-ABC-modulation}
and \eqref{eq:expected-mod}, in the small-angle limits where those equations
hold.
As the opening angle of the emitted pair departs from that limit, the
showers' azimuthal modulations deform from a pure cosine structure,
as will be seen later, cf.\ the red dashed lines of
Fig.~\ref{fig:coll-spin-as2-issue} in Sec.~\ref{sec:me-spin}. 
Analytically computing the effective parton shower double-soft matrix
elements, including spin correlation effects, gives expressions of the form,
\begin{align}
  \label{eq:spin-dependent-omega}
  \omega^{(S)}_{\mathrm{spin}} \, = \, \sum_h\omega^{(S)}_h \left[1+\mathcal B_{kg}(z_{2,h}){\mathcal{C}}_{h}(\phi)\right]\,.
\end{align}
In the double-soft limit, the $\mathcal{C}_{h}(\phi)$ factors are somewhat
cumbersome functions of the six Lorentz invariants that can be formed
from the momenta of the hard antenna and the emitted soft pair.
They encode not only the physical azimuthal modulation at small
angles, but also its non-trivial deformations away from that limit.
It is tempting to think that one might promote the double-soft matrix 
element correction procedure to work consistently with spin correlations
by replacing each $\omega^{(S)}_h$ by the corresponding spin-dependent
summand in Eq.~\eqref{eq:spin-dependent-omega}.
Such a replacement is too optimistic for the following reason.
With spin correlations enabled, the PanScales accept--reject loop in
the veto algorithm repeatedly generates $\phi$ values at \emph{fixed}
pre-branching kinematics, $\ln v$, $\bar\eta$, until one is eventually
accepted.
That gives an effective matrix element that involves
each term in the sum of
Eq.~(\ref{eq:spin-dependent-omega}) being divided by its average over
$\phi$.
The key issue is that outside the collinear limit, the complexity of
the ${\mathcal{C}}_{h}(\phi)$ prevents us from straightforwardly
determining that average.
This makes it complicated to generalise the $\omega^{(S)}_h$ shower
weights to include spin correlations.

The solution we adopt is to modify the PanScales spin-correlation
algorithm.
Instead of using Eqs.~\eqref{eq:amp-old}--\eqref{eq:Sbb-def}, we
always take their collinear limit, even when emissions are at
large angles.
The collinear limit depends on the azimuth only through the orientation of $k_\perp$, and so
guarantees a pure cosine modulation for arbitrary opening angles.
This is legitimate because the difference between $\mathcal{C}_h(\phi)$ and
$\cos(2\Delta\psi_{12})$ resides entirely at large angles, which is a
region that is anyway beyond our spin-correlation accuracy.

The specific limit we take is the one in which the radiation becomes collinear to the
splitter, at fixed momentum fraction $z$, so that it acts on the spinor structure alone and
leaves the momentum-fraction dependence untouched. Taking this limit in Eq.~\eqref{eq:amp-old},
with $i$ as the splitter, gives
\begin{align}
\label{eq:coll-limits}
S_{\lambda}(i,k) \; \to \;
\sqrt{\,2\, i\cdot k} \; \frac{ {\mathbb  {S}}_{-\lambda}(\tilde{\imath},{\widehat k}_\perp) }{\tilde{\imath}\cdot k_{0}}  \, ,
\qquad
\frac{S_{\lambda}(i,j)}{S_{\lambda}(k,j)} \; \to \; \sqrt{\frac{z}{1-z}} \, ,
\end{align}
where ${{\widehat k}_{\perp}}     = n_1 \, \cos\phi\, + n_2 \,
\sin\phi$, and the other-helicity
spinors are obtained as usual by $ S_{-}(a,b) = - S_{+}^{*}(a,b)$.
The soft correction factor in Eq.~\eqref{eq:amp-old}, in this limit,
becomes one, and the helicity amplitude reduces to
\begin{align}
\label{eq:amp-new}
   A^{\lambda\,\lambda_1 \lambda_2} \, = \,
     - \sqrt{2} \, g_{s} \;
       \mathcal{F}^{\lambda\lambda_1\lambda_2}(z) \;
       \frac{1}{\sqrt{\,2\, i\cdot k}} \;
       \frac{\tilde{\imath}\cdot k_{0}} { {\mathbb  {S}}_{\tau}(\tilde{\imath},{\widehat k}_\perp) }
       \, .
\end{align}
To see the $\phi$-independence of the branching probability in the soft limit, it may help to rewrite it as
\begin{align}
\label{eq:amp-new-KH}
  A^{\lambda\,\lambda_1\lambda_2}  
  \, = \, -\sqrt{2} \, g_s \, \mathcal{F}^{\lambda\lambda_1\lambda_2}(z) 
       \, \frac{1}{\sqrt{2\,i\cdot k}} \, U_{\tau}(\phi) \,,
  \quad
  U_{\tau}(\phi) 
  \, = \, \frac{\tilde{\imath}\cdot k_0}{\mathbb{S}_{\tau}(\tilde{\imath},\widehat{k}_{\perp})} \,,
  \quad
  |U_{\tau}(\phi)|^{2} \, = \, 1 \,,
\end{align}
free of any reference to the spectator. We use this form, 
which we call \emph{collinear spin projection}, to generate
the azimuthal angle $\phi$ of the branching in the spin-correlation acceptance loop.
The azimuth $\phi$ enters Eq.~\eqref{eq:amp-new} only through $\hat k_\perp$, so the amplitude generates a pure cosine modulation
at any opening angle, and the collinear projection
leaves the modulation in collinear splittings untouched.

The collinear projections above are only used in the spin-dependent azimuthal
acceptance probability when \emph{generating} the $\phi$ angle of a branching. 
Once a branching is accepted, the original amplitude, Eq.~\eqref{eq:amp-old},
evaluated with physical momenta, is still used to update all internal spin-density
and decay matrices.
In other words, only the branching amplitudes for the \emph{generation} of
$\phi$ are modified relative to the existing PanScales
spin-correlation algorithms, with the definition and upkeep of all spin-densities
wholly unchanged.
This minor, surgical, intervention does not compromise the algorithms'
SL accuracy.\footnote{We have explicitly checked numerically that the
  PanScales showers with collinear-spin projection pass the all-order
  logarithmic accuracy tests of
  Refs.~\cite{Karlberg:2021kwr,Hamilton:2021dyz}.}
In the collinear limit, the original branching amplitudes and their collinear
projections --- Eqs.~\eqref{eq:amp-old} and \eqref{eq:amp-new} respectively ---
lead to the same spin-dependent $\phi$ distributions by construction.
In the soft limit, regardless of angle, both sets of branching amplitudes
conserve the emitter's helicity, with non-trivial helicity dependence limited
to the emitted gluon's polarisation ($\lambda_{2}$). The helicity indices of
the parent dipole particles can then be trivially contracted away in constructing
the density matrices, with further sums over the emitted gluon's helicity
index yielding $\phi$-independent soft-gluon emission probabilities.\footnote{In the
soft limit, Eq.~\eqref{eq:amp-old} contracts to yield the familiar eikonal
dipole factor, cf. Eqs.~(2.4)--(2.5) of Ref.~\cite{Hamilton:2021dyz}.} 
At amplitude level, emitted soft gluons nevertheless encode information about
their emitting antenna in their own polarisation, which the branching amplitudes
of Ref.~\cite{Hamilton:2021dyz}, evaluated on physical momenta, take full account
of, regardless of angle. Since the modified spin-correlation algorithm still
employs all the same branching amplitudes as Ref.~\cite{Hamilton:2021dyz} to
update the underlying density matrices, showers initiated from soft wide-angle
gluons propagate physical azimuthal correlations between themselves
and the original parent antenna as before.
The generation prescriptions therefore agree in the soft and strongly ordered
collinear limits, while the amplitudes used for spin propagation remain unchanged.
Differences between the prescriptions are therefore beyond SL accuracy.

With these adaptations of the spin correlation algorithms in
effect, we repeat the derivation of the shower's effective matrix
element for the double-soft correction denominator,
$\omega^{(S)}_\mathrm{spin}$, taking them into account. As expected, 
that matrix element acquires only a multiplicative factor compared to
the spin-averaged one,
\begin{align}
\label{eq:M-spin-final}
\omega^{(S)}_\mathrm{spin} = \sum_h\omega^{(S)}_h\,
\big[\,1 + \mathcal{B}_{kg}(z_{2,h})\,\mathcal{C}_h(\phi)\big]\,,
\end{align}
with $\mathcal{B}_{kg}(z)$ given in Eq.~\eqref{eq:Bkg}. The arguments of $\mathcal{B}_{kg}$
and the $\mathcal{C}_h(\phi)$ are shower-, and history-,
dependent, and are
given in Appendix~\ref{app:spin-corr-Cab}, as functions of the
post-branching invariants.
By construction, the average of the square bracket in Eq.~\eqref{eq:M-spin-final}
over the azimuthal angle $\phi$ of the second emission is equal to 1.

We implement the above in the PanScales code simply by dressing the
spin-independent shower double-soft splitting weights with the factor
in the bracket of Eq.~\eqref{eq:M-spin-final} whenever the
(collinear-projected) spin correlations are enabled.
We note that with strong angular ordering the double-soft corrections
become $1$ which ensures that we retain the correct SL spin
correlations in the collinear limit.

\section{Validation and timings}\label{sec:analytic-tests}

In this section we provide a validation of our approach.
In particular, first we perform fixed-order tests which illustrate how we reproduce the correct matrix element for a pair of emissions~(\ref{sec:phase-space} on
the full coverage of phase space, \ref{sec:me-double-soft} on comparisons to the exact double-soft matrix element),
as well as tests of the collinear spin projection~(\ref{sec:me-spin}).
We then perform all-order numerical validations to demonstrate NNDL accuracy for
subjet multiplicities~(\ref{sec:multiplicity}), NSL accuracy
for non-global observables~(\ref{sec:non-global}), and finally NNLL
accuracy for global event shapes~(\ref{sec:nnll-shapes}).

We remind the reader that the analytic proof
presented in Ref.~\cite{vanBeekveld:2024wws} is independent of the particular
NLL shower used as a starting point, and therefore remains valid also in this case.

\subsection{Fixed-order tests}
\label{sec:fo-tests}

\subsubsection{Phase-space coverage}\label{sec:phase-space}

First, we illustrate that disordered emissions enable us to obtain
bounded double-soft overheads for the \pl{} shower,
and substantially reduce the required double-soft overhead in the \pg{} case.

\begin{figure}
  \centering
  \begin{subfigure}[t]{0.48\textwidth}
    \centering
    \includegraphics[width=\textwidth]{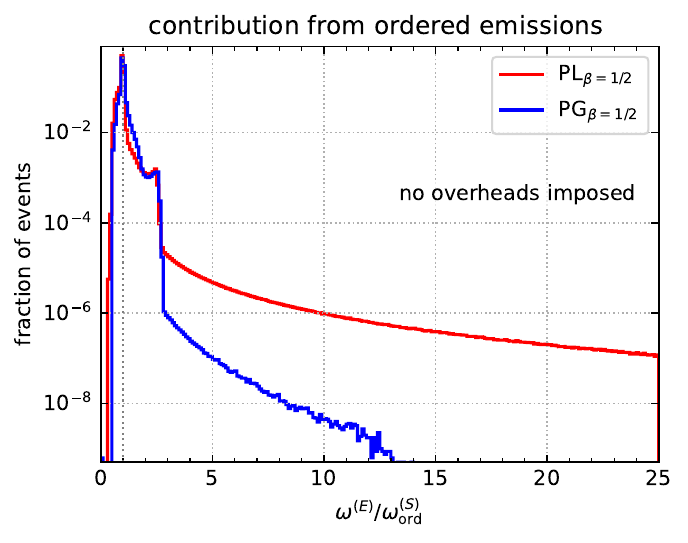}
    \caption{Ratio between exact ($\omega^{(E)}$) and shower
      double-soft effective ME 
      considering only ordered emissions  ($\omega^{(S)}_\text{ord}$).
    }
    \label{fig:phsp1}
  \end{subfigure}
  \hfill
  \begin{subfigure}[t]{0.48\textwidth}
    \centering
    \includegraphics[width=\textwidth,page=3]{figs/test-coverage.pdf}
    \caption{Ratio between the remnant of the exact double-soft
      weight after ordered emissions
      ($\omega^{(E)} - \Omega \omega^{(S)}_\text{ord}$) and the shower's
      effective weight for disordered configurations ($\omega^{(S)}_\text{disord}$).
    }
    \label{fig:phsp2}
  \end{subfigure}
  \caption{Distribution of double-soft reweighting factors, without
    (a) and
    with (b) disordering, considering the  \pg~(blue) and \pl~(red)
    showers, with $\betaps{}=\frac{1}{2}$.
  }
  \label{fig:phspall}
\end{figure}

To test the phase-space coverage of the parton showers in the
double-soft limit, we have written an independent phase-space
generator\footnote{
  In the double-soft limit, the phase-space only depends on the
  directions of the four partons and the relative energy fraction of
  the double-soft pair.
  In practice, we align one of the hard partons with the $z$ axis and
  for each of the other three partons, sample
  $\frac{1}{2}\ln\frac{2}{1-\cos\theta}$ between 0 and some
  $\eta_\text{max}$ and their azimuth $\phi$, setting $\phi=0$ for
  the second hard parton.
}
and checked that (a)~each phase-space point was accessible
through at least one (potentially disordered) shower history and
(b)~the full QCD matrix element could be reproduced with a given
finite overhead.
This latter point is illustrated in Fig.~\ref{fig:phsp1}.
We take our NLL showers with $\betaps{}=0.5$ and plot the
distribution of the ratio between the predictions of
$\omega^{(S)}_{\rm ord}$ and the exact double soft matrix element
$\omega^{(E)}$.
As expected, for the majority of the events this ratio is 1,
a consequence of the fact that our showers need to be corrected only for two
neighbouring emissions.
We can see however that both showers have substantial tails.
In the case of the PanGlobal shower, the overhead of $20$ that we took
in Ref.~\cite{FerrarioRavasio:2023kyg} can be seen to provide ample
margin.\footnote{Some of this margin is, we believe, taken up by the
  fact that we apply double-soft corrections outside the pure
  double-soft kinematic limit, which can lead to larger ratios than
  appear in Fig.~\ref{fig:phsp1}.}
For the PanLocal shower, there is a very long tail.
While the ordered shower covers phase space, the ordered acceptance
ratio, Eq.~(\ref{eq:Pacc-ordered-only}), can grow arbitrarily
large, as we saw also in Fig.~\ref{fig:PE-MS-ratio}.

If we instead allow for disordered emissions, then we can
reproduce the exact double-soft matrix element while using a much
more contained overhead, for example $\Omega=2$ for \pg, and
$\Omega=3$ for \pl.
The disordered contributions are responsible for the leftover contribution,
\begin{equation}
  \Delta \omega = \max\left(\omega^{(E)} - \Omega \omega^{(S)}_{\rm ord},\,0\right).
\end{equation}
In Fig.~\ref{fig:phsp2} we show the ratio between $\Delta \omega$ and the shower effective ME only considering disordered emissions, $\omega^{(S)}_{\rm disord}$, with $\ln v_1 < \ln v_2 < \ln v_1+\Delta$ with $\Delta=1$.
We see that this contribution is very small and in particular the
ratio with $\Omega \omega^{(S)}_{\rm
  disord}$, Eq.~(\ref{eq:disordered-Pacc-disordered}), 
will be smaller than one (note the figure does not include the factor
of $\Omega$ in the denominator).
For simplicity, we apply the same overhead $\Omega$ to the ordered and
disordered cases.


\subsubsection{Double-soft matrix elements with disordered emissions}
\label{sec:me-double-soft}

\begin{figure}
  \begin{subfigure}[c]{0.45\textwidth}
    \includegraphics[width=\textwidth,page=1]{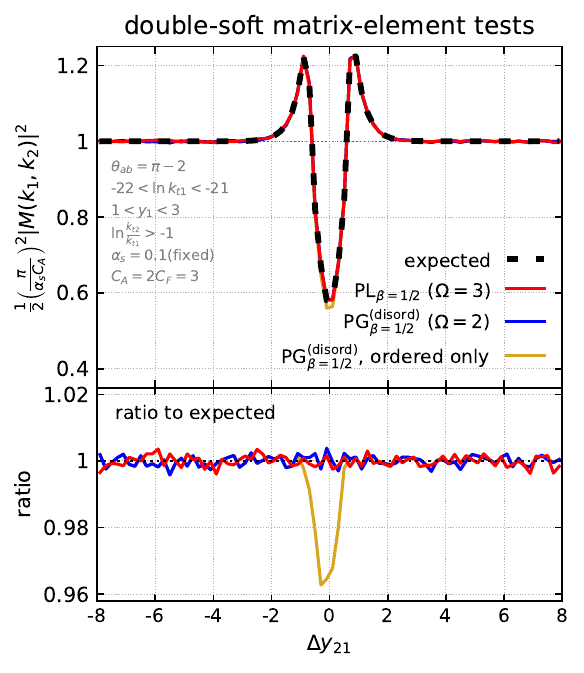}
    \caption{Shower double-soft matrix element compared to the
      QCD expectation. }
    \label{fig:me-tests-full}
  \end{subfigure}
  \hfill
  \begin{subfigure}[c]{0.53\textwidth}
    \includegraphics[width=\textwidth,page=2]{figs/me-tests-disordered.pdf}
    \caption{Shower double-soft matrix element separated in flavour channels
      and colour flows.}
    \label{fig:me-tests-channels}
  \end{subfigure}
  \caption{Double-soft matrix-element tests at
    $\mathcal{O}(\alpha_s^2)$ for PanLocal($\betaps=\tfrac{1}{2}$), in red, and PanGlobal($\beta=\tfrac{1}{2}$), in blue.
		Both showers include disordered emissions.
	In the left plot, we also show the contribution of only ordered emissions for PanGlobal($\beta=\tfrac{1}{2}$), while keeping the same ordered
        overhead of $\Omega=2$ (yellow).
      }
  \label{fig:me-tests}
\end{figure}

In this section we show one of
the matrix-element tests that we have carried out to verify that we correctly describe the production of any pair of soft emissions.
We start from a dipole $ab$ with an opening angle $\theta_{ab}=\pi-2$ in the event centre-of-mass frame.
We then generate two soft emissions across the whole phase space
and
we select events where the emission with the largest reconstructed
Lund $k_t$ satisfies $-22
< \ln k_{t1}/Q < -21$ and $1 < y_1 < 3$, while the other satisfies $\ln k_{t2}/k_{t1} > -1$.
Rapidities and transverse momenta are computed in the dipole centre-of-mass frame and we consider only cases in which both emissions lie in the $ab$ dipole's primary Lund plane~\cite{Dreyer:2018nbf}.
We use our soft phase space generator from the previous section
together with the true double-soft matrix element, binned as a
function of $\Delta y_{21}$ (``expected''); and we use the shower with
its phase space generation, including disordering and the double-soft
correction, to get the corrected shower matrix element. In
Fig.~\ref{fig:me-tests-full} we compare the two for the
\pg{} ($\betaps=1/2$) shower (in blue, with an overhead
$\Omega=2$) and for the \pl ($\betaps=1/2$) shower (red, $\Omega = 3$).
Both use disordered emissions.
We also show the contribution given by only ordered emissions, focusing
on the \pg{} shower (yellow), still with $\Omega=2$ (we remind the
reader that to have acceptance probabilities below $1$ with just
ordered emissions, one would need
$\Omega\approx20$~\cite{FerrarioRavasio:2023kyg}).

The upper panel shows the azimuthally-averaged differential distribution of the rapidity
difference between the two emissions, $\Delta y_{21}=y_2-y_1$.
The results have been normalised to
$2\,\!(\as C_A/\pi)^2$, which corresponds to the non-correlated contribution to the double-soft analytic prediction.
In the bottom panel, we show the ratio with the expected analytic result.
We begin by noticing that, with the inclusion of disordered emissions,
both \pg{} and \pl{} agree with the analytic expectation.
The discrepancy between the ordered-only~(yellow) contribution of
\pg{} and the full result appears only for small values of $|\Delta
y_{21}|$, confirming that the insertion of disordered emissions is
localised to an $\mathcal{O}(1)$ rapidity region, which does not interfere with the overall NLL accuracy of the starting shower algorithm.

In Fig.~\ref{fig:me-tests-channels}, we show the same distribution, separated into colour and flavour channels, focusing on the showers that include disordered emissions. We see that also in this case the parton-shower predictions agree with the analytic expectation.
We stress that the correct modelling of the colour flows and of the parton flavours is paramount for non-global and multiplicity-like observables.

\subsubsection{Spin correlations}\label{sec:me-spin}

First, we illustrate the issue with our previous default implementation of spin correlations at
large angle, and the impact of the new collinear spin projection of Eqs.~\eqref{eq:coll-limits},~\eqref{eq:amp-new}
at fixed order. In order to do so, we generate a soft gluon at central rapidity
($\ln k_{t1}/Q \lesssim -7$, and $y_1 = 0.4$); we force this gluon to split at a given $z_2 = 0.385$,
either collinearly ($\bar \eta_2 = 8.0$), or at large angle ($\bar \eta_2 = 0.2$).
We then extract the modulation of the shower weight stemming from the spin-correlation algorithm
in the acceptance loop for the azimuth of the second splitting,
i.e.\ the dependence on the shower-generation variable
$\phi_2$. 

\begin{figure}
  \begin{subfigure}[c]{0.49\textwidth}
    \centering
    \includegraphics[width=\textwidth,page=1]{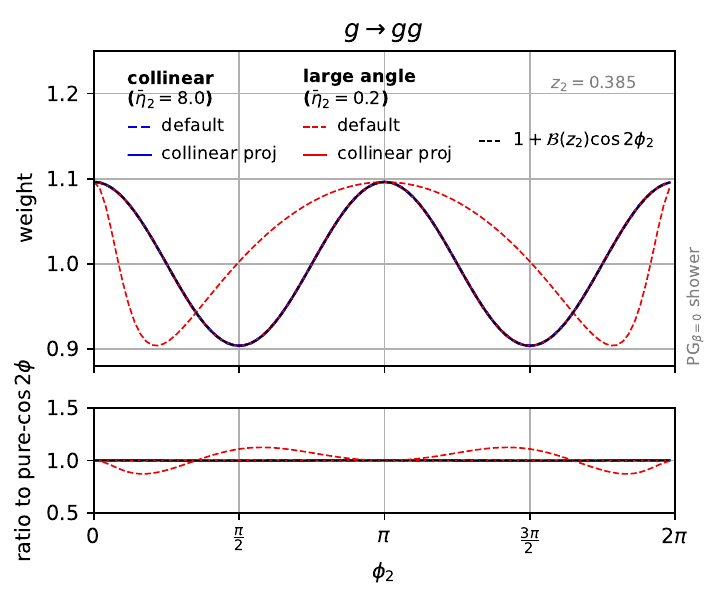}
    \caption{$g\to gg$ splitting.}
  \end{subfigure}
  \begin{subfigure}[c]{0.49\textwidth}
    \centering
    \includegraphics[width=\textwidth,page=2]{figs/plot-coll-spin.pdf}
    \caption{$g\to q\bar q$ splitting.}
  \end{subfigure}
  \caption{
    Second-order tests of the shower's spin-correlation algorithm for (a) $g\to gg$ and (b) $g\to q\bar{q}$ splittings.
    The default implementation of spin correlations
    (dashed) fails to produce the pure $\cos2\phi$ modulation at large angles (red).
    In all other cases (collinear configurations, or when replacing the default amplitudes with the spin-projection version, solid),
    all curves agree with the pure $\cos2\phi$ expectation.
  }
  \label{fig:coll-spin-as2-issue}
\end{figure}

This is shown in Fig.~\ref{fig:coll-spin-as2-issue}, for $g\to gg$ (left) and $g\to q\bar q$ (right).
We compare the shower weight to the pure $\cos2\phi$ modulation, with $\mathcal{B}(z_2 = 0.385)$ from Eq.~\eqref{eq:Bkg} (dashed black).
In collinear splittings (blue), the shower reproduces the pure $\cos2\phi$ modulation, whether with
the old default (dashed) or the new collinear spin projection (solid). However, at large angles (red),
the default implementation (dashed red) produces a distorted modulation; only when the collinear
projection is applied (solid red) does the shower reproduce the pure $\cos2\phi$ modulation. We have
checked that this is true for any value of $z_2$ and arbitrary opening angles in the splitting of the
soft gluon $g_1$.

Next we test the shower's double-soft radiation pattern, including the collinear
spin projections presented in Sec.~\ref{sec:ds-and-spin}, in the double-soft region of phase space.
We generate a pair of soft partons ($gg$ or $q\bar q$) and perform a Lund
declustering to identify the splittings. As in the last section, we require the first, harder
splitting to have transverse momentum $-12 < \ln k_{t1} < -11$ and $1 < y_1 < 3$,
and the second, softer splitting to lie within one
logarithmic unit, $\ln k_{t2}/k_{t1} > -1$.

\begin{figure}
  \begin{subfigure}[c]{0.49\textwidth}
    \centering
    \includegraphics[width=\textwidth,page=1]{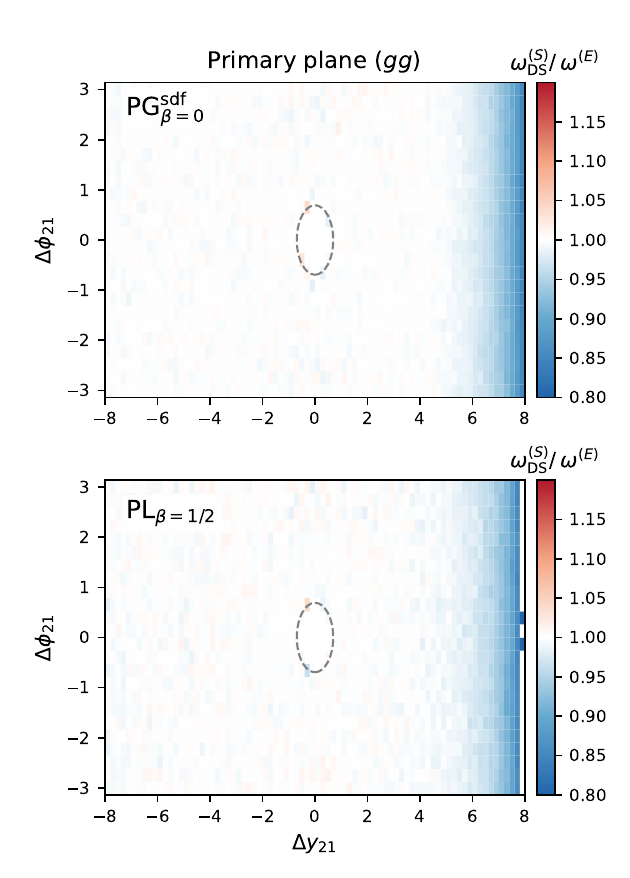}
    \caption{$g\to gg$ splitting.}
  \end{subfigure}
  \begin{subfigure}[c]{0.49\textwidth}
    \centering
    \includegraphics[width=\textwidth,page=2]{figs/plot-double-soft-w-spin.pdf}
    \caption{$g\to q\bar q$ splitting.}
  \end{subfigure}
  \caption{
    The ratio between the binned shower event weights, $\omega^{(S)}_{\mathrm{spin,DS}}$,
    with collinear spin and double-soft corrections, and
    the binned exact double-soft matrix element,
    $\omega^{(E)}$ for \pgzerosdf (top row) and \plhalf (bottom row). 
    The binning is in $\Delta y_{21}$ and $\Delta \phi_{21}$,
    with further cuts as described in the main text.
    At $\Delta y_{21} \gtrsim 6$, the second emission
    becomes hard and collinear to the quark, and the
    shower's effective matrix element departs from the
    analytic (double-soft) result $\omega^{(E)}$,
    irrespective of the spin modulation.
  }\label{fig:spin-tests-ratio}
\end{figure}

In Fig.~\ref{fig:spin-tests-ratio}, we show the ratio between the
shower's binned matrix element with collinear spin and double-soft
corrections enabled, $\omega^{(S)}_{\mathrm{spin,DS}}$, and the binned
exact double-soft matrix element, $\omega^{(E)}$.
This figure is effectively a 2D analogue of the ratio panel of
Fig.~\ref{fig:me-tests-full}, keeping in mind that
Fig.~\ref{fig:me-tests-full} did not include spin correlations.
The shower and analytic results are in good agreement, confirming that
the shower correctly reproduces the expected double-soft 
radiation pattern.
In the region where the second emission becomes hard and collinear to the quark ($\Delta y_{21} \gtrsim 6$ on the left-hand figure),
the shower's effective matrix element is not expected to reproduce the analytic result, which was
derived in the double-soft limit. Note also that while we do not show the secondary Lund plane ---
because the phase space is very limited by the selection $\ln k_{t2}/k_{t1} > -1$ --- the predictions
are also in perfect agreement when the two soft particles are collinear to each
other.
There is qualitatively the same agreement for all the showers
considered in this work.\logbook{}{See other plots in the same
  directory as the one displayed}

\subsection{All-order validation}
\label{sec:ao-tests}
We now perform all-order tests of the logarithmic accuracy of
our showers.
As in Ref.~\cite{Dasgupta:2020fwr} and subsequent work, we run
the shower for several small but finite values of $\as$ (with fixed
$\as L$ or $\as L^2$) and then fit a polynomial to those results to
carry out an $\as \to 0$ extrapolation.
That extrapolation is intended to eliminate contributions that are
subleading with respect to the logarithmic order being tested, so that
one can reliably compare to analytic predictions.
One significant issue that we had faced in that approach was knowing
how to choose the degree of the polynomial: too low a degree can
result in a biased answer, notably in the presence of numerically
relevant higher powers of $\as$; however, too high a degree
leads to overfitting and artificially enlarged errors.
For the results that we present below, we address the issue by
adopting the Akaike information criterion
(AIC)~\cite{Akaike:1974vps}.
In essence, we carry out fits with multiple choices of polynomial
degree.
The AIC approach then weights different degrees based on the goodness
of fit and an additional penalty term for higher degrees.
In practice the final extrapolation result is usually dominated by the
lowest degree polynomial that is consistent with the data to within
its statistical errors.
Appendix~\ref{app:extrapolation-strategy} provides further details,
including considerations for fits with a small number of $\as$ points,
the treatment of errors, and illustrations with specific fits.

The comparisons shown below are all performed in the large $N_c$
limit, $C_A = 2C_F = 3$, with $n_f=5$ light flavours.
We will explore a range of shower configurations, including:
\begin{compactitem}
  \item the \pg{} shower with $\beta_{\ps}=0$, with $\bar \eta$ defined in
    the dipole frame (``sdf''~\cite{FerrarioRavasio:2023kyg}), without
    spin correlations, $\text{PG}_{\beta=0}^\text{(sdf)}$ (this is the
    strictly ordered variant, as used in Ref.~\cite{vanBeekveld:2024wws});
  \item the \pg{} shower with $\beta_{\ps}=1/2$, including disordered
    emissions, without spin correlations, $\text{PG}_{\beta=1/2}^\text{(disord)}$;
  \item the \pl{} shower with $\beta_{\ps}=1/2$, including disordered
    emissions, without spin correlations, $\text{PL}_{\beta=1/2}$;
  \item the \pl{} shower with $\beta_{\ps}=1/2$, including disordered
    emissions, with spin correlations, $\text{PL}_{\beta=1/2}^\text{(spin)}$.
\end{compactitem}
Note that only the last of these includes spin correlations.
This is related to the fact that none of the NNDL/NSL/NNLL
tests shown below directly probe the full nested spin-correlation
tree.
There is however an indirect effect of spin correlations in that it
changes the shower effective matrix elements and hence the drifts (see
Table~\ref{tab:drifts}) and associated $\Delta K(\bar \eta)$ function
(see Eq.~(\ref{eq:aseff})).
Therefore we felt it was important to include at least one shower with
spin correlations turned on, which is  $\text{PL}_{\beta=1/2}^\text{(spin)}$.

\subsubsection{Subjet multiplicity at NNDL accuracy}\label{sec:multiplicity}
The first observable we examine is the Lund subjet multiplicity in
lepton collisions~\cite{Medves:2022ccw}.
This observable has a double-logarithmic expansion, whose leading
terms scale as $\alpha_s^n L^{2n}$.
The original PanScales showers of Ref.~\cite{Dasgupta:2020fwr} already
described it at NDL accuracy, including terms of order
$\alpha_s^n L^{2n-1}$.
Including matching corrections~\cite{Hamilton:2023dwb} and the
double-soft matrix elements raises the logarithmic accuracy of the
shower to NNDL~\cite{FerrarioRavasio:2023kyg}, i.e.~up to terms of order
$\alpha_s^n L^{2n-2}$.
Here we explicitly verify that this remains the case when disordered
emissions are included.
To do so, we test whether the following quantity vanishes:
\begin{equation}
  \label{eq:NNDL-test}
  \Delta^\text{(Lund)}_{\textsc{nndl}} = \lim_{\sqrt{\as} \to 0}
  \left. \frac{N_\ps - N_\nndl}{\as N_\dl}
  \right|_{\text{fixed $\xi=\as L^2$}},
  \qquad L \equiv \ln \frac{k_{t,\min}}{Q},
\end{equation}
where $N_\ps$ is the parton-shower result, $N_\nndl$ and $N_\dl$ are
the known analytic NNDL and DL results~\cite{Medves:2022ccw},
respectively, $Q$ is the centre-of-mass energy, and $k_{t,\min}$ is
the subjet-resolution scale.
The extrapolation $\as\to 0$ is performed numerically using the new
AIC strategy based on $\as=$ 0.0001953125, 0.00032, 0.0005,
0.00078125, 0.00128, 0.002, 0.003125, 0.00512, 0.008, 0.0125
($\text{PL}_{\beta=1/2}$ in red, and
$\text{PG}_{\beta=1/2}^\text{(disord)}$ in blue), or a simple cubic
polynomial extrapolation based on $\as=$ 0.0001953125, 0.002,
0.003125, 0.008 (our previous result for
$\text{PG}_{\beta=0}^\text{(sdf)}$ in green, and a lower statistics
$\text{PL}_{\beta=1/2}$ run without double-soft corrections, in yellow).
In both cases the polynomials are in $\sqrt{\as}$, as is appropriate
given the structure of the double-logarithmic hierarchy at fixed
$\as L^2$.

\begin{figure}
  \centering
  \includegraphics[width=0.6\textwidth]{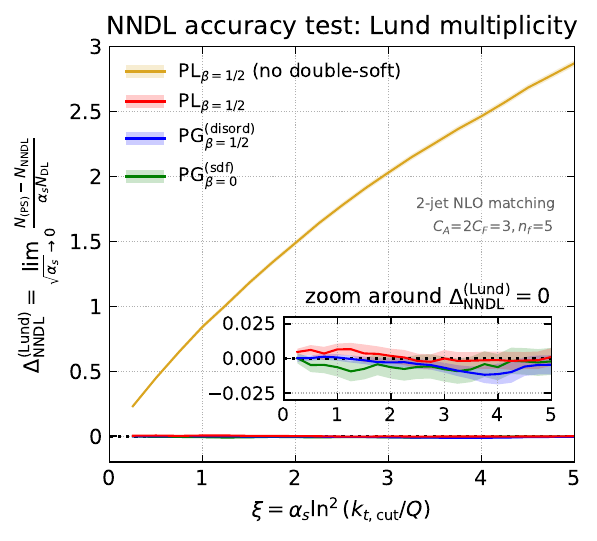}
  \caption{NNDL test for the Lund subjet multiplicity for several showers. The quality
    measure $\Delta^\text{(Lund)}_{\textsc{nndl}}$, defined in
    Eq.~\eqref{eq:NNDL-test}, should vanish for an NNDL-accurate
    shower for all values of $\xi=\alpha_s L^2$.
    The inset shows a magnified view around $\Delta=0$.}
  \label{fig:nndl-multiplicity}
\end{figure}        
The results are shown in Fig.~\ref{fig:nndl-multiplicity}.
As expected, $\Delta^\text{(Lund)}_{\textsc{nndl}}$ is consistent with zero
for the showers that include double-soft corrections, but not for the
original \pl{} shower without those corrections.

\subsubsection{Non-global logarithms at NSL accuracy}
\label{sec:non-global}

\begin{figure}
  \centering
  \includegraphics[width=0.6\textwidth]{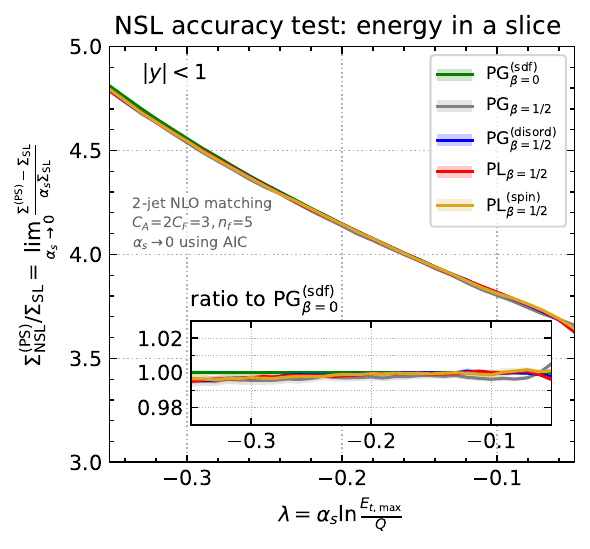}
  \caption{NSL test for the energy flow in a rapidity slice. The
    main panel shows the normalised NSL contribution for several
    PanScales showers. The
    inset shows the ratio to the PG$_{\beta=0}^\text{(sdf)}$ shower
    (which, we recall, is without disordered emissions).}
  \label{fig:nsl-energy-slice}
\end{figure}

We now turn to non-global observables, sensitive to large-angle soft
emissions. Compared with the subjet multiplicity, their NSL
description additionally requires the soft wide-angle NLO correction
$\Delta K_1(\bar \eta)$, as used in Eq.~(\ref{eq:aseff}), including
its fully differential kinematic dependence on the parent dipole
kinematics. 

We consider the transverse energy flow in a rapidity slice, defined as the sum of
the transverse energies $E_t$ of all particles in the slice.
We choose the slice to involve rapidities $|y|<1$, and
rapidity is measured with respect to an axis obtained by clustering
the event into two jets with the Cambridge
algorithm~\cite{Dokshitzer:1997in}.
The single-logarithmic (SL) terms scale as
$\mathcal{O}(\alpha_s^n L^n)$, while the next-to-single-logarithmic
(NSL) corrections scale as $\mathcal{O}(\alpha_s^{n+1}L^n)$.
We denote by $\Sigma$ the cumulative distribution obtained by
requiring the summed transverse energy to be below $E_{t,\max}$.
Analytic NSL predictions for $\Sigma$ have been computed in
Refs.~\cite{Banfi:2021owj,Banfi:2021xzn,Becher:2021urs,Becher:2023vrh,Becher:2026zon}.
Comparisons to Ref.~\cite{Banfi:2021xzn}  were carried out
in Ref.~\cite{FerrarioRavasio:2023kyg} for $n_f=0$ in the real
emission (but full $n_f$ in the running of the coupling, etc.).
Further comparisons, including $n_f=5$ results, are ongoing~\cite{NGLComparison}.

Fig.~\ref{fig:nsl-energy-slice} shows the normalised NSL
parton-shower contribution,
\begin{equation}
  \label{eq:nsl-log-test}
  \frac{\Sigma_{\nsl}^{(\ps)}}{\Sigma_\SL}
  = \lim_{\as \to 0}
  \left. \frac{\Sigma^{(\ps)}-\Sigma_\SL}{\as\Sigma_\SL}
  \right|_{\text{fixed $\lambda=\as L$}},
  \qquad L \equiv \ln\frac{E_{t,\max}}{Q},
\end{equation}
where $\Sigma_\SL$ is the numerical SL prediction.
The extrapolation $\as\to 0$ is performed using the AIC strategy based
on $\as=$ 0.002, 0.0025, 0.0032, 0.004, 0.005, 0.0064, 0.008, 0.01,
0.0125, 0.016, 0.02.
The \pg{} predictions with $\betaps=0$ in the split-dipole frame (green)
and with $\betaps=1/2$ without disordered emissions (grey) are equivalent to those of
Ref.~\cite{FerrarioRavasio:2023kyg}, which were tested against
\textsc{Gnole}~\cite{Banfi:2021xzn} for $n_f=0$.
The new \pl{} and \pg{} showers with $\betaps=1/2$ and disordered
emissions (red and blue, respectively) agree with the previously
validated NSL-accurate \pg{} predictions.
This agreement provides a
numerical validation of NSL accuracy for the new showers.
Since there is no purely analytic result for the NSL resummation,
Table~\ref{tab:ngl-results} shows the numerical values associated with
Fig.~\ref{fig:nsl-energy-slice}.
We hope that this may facilitate comparisons with other groups'
results.

\begin{table}
  \centering
  \footnotesize
  \begin{tabular}{cccccccc}
    \toprule
$\lambda$     & PG$_{\beta=0}^{(\mathrm{sdf})}$  & PG$_{\beta=0}$  & PG$_{\beta=\frac{1}{2}}$  & PG$_{\beta=\frac{1}{2}}^{{\rm (disord)}}$  & PL$_{\beta=\frac{1}{2}}$  & PL$^{\rm (spin)}_{\beta=\frac{1}{2}}$  & PL$_{\beta=\frac{1}{2}}^{{\rm (min)}}$ \\
\midrule
$ -0.35$  &  $4.810(7)$ &$4.817(10)$ & $4.787(14)$ & $4.789(6)$ & $4.787(8)$ & $4.794(11)$ & $4.795(13)$  \\
$ -0.34$  &  $4.756(7)$ &$4.765(9)$ & $4.734(14)$ & $4.735(6)$ & $4.737(8)$ & $4.740(11)$ & $4.742(12)$  \\
$ -0.33$  &  $4.703(6)$ &$4.710(9)$ & $4.681(14)$ & $4.683(6)$ & $4.686(7)$ & $4.684(11)$ & $4.689(11)$  \\
$ -0.32$  &  $4.652(6)$ &$4.659(9)$ & $4.636(13)$ & $4.638(6)$ & $4.636(6)$ & $4.637(11)$ & $4.640(10)$  \\
$ -0.31$  &  $4.605(6)$ &$4.609(8)$ & $4.586(13)$ & $4.590(5)$ & $4.589(6)$ & $4.590(11)$ & $4.596(9)$  \\
$ -0.30$  &  $4.557(5)$ &$4.559(8)$ & $4.538(12)$ & $4.545(5)$ & $4.544(7)$ & $4.545(11)$ & $4.550(8)$  \\
$ -0.29$  &  $4.510(5)$ &$4.512(8)$ & $4.491(12)$ & $4.497(5)$ & $4.501(7)$ & $4.496(10)$ & $4.506(8)$  \\
$ -0.28$  &  $4.467(5)$ &$4.468(8)$ & $4.448(11)$ & $4.456(5)$ & $4.456(6)$ & $4.453(9)$ & $4.461(8)$  \\
$ -0.27$  &  $4.422(5)$ &$4.426(7)$ & $4.407(11)$ & $4.416(5)$ & $4.415(5)$ & $4.412(9)$ & $4.419(8)$  \\
$ -0.26$  &  $4.379(4)$ &$4.379(7)$ & $4.363(11)$ & $4.375(4)$ & $4.372(5)$ & $4.370(8)$ & $4.375(7)$  \\
$ -0.25$  &  $4.339(4)$ &$4.337(7)$ & $4.325(10)$ & $4.333(4)$ & $4.331(4)$ & $4.330(8)$ & $4.333(7)$  \\
$ -0.24$  &  $4.300(5)$ &$4.292(7)$ & $4.292(10)$ & $4.295(4)$ & $4.292(4)$ & $4.291(8)$ & $4.295(7)$  \\
$ -0.23$  &  $4.261(4)$ &$4.254(6)$ & $4.250(10)$ & $4.255(4)$ & $4.253(3)$ & $4.256(8)$ & $4.258(6)$  \\
$ -0.22$  &  $4.222(4)$ &$4.216(6)$ & $4.210(9)$ & $4.218(4)$ & $4.216(3)$ & $4.218(7)$ & $4.219(6)$  \\
$ -0.21$  &  $4.185(4)$ &$4.179(6)$ & $4.174(9)$ & $4.182(4)$ & $4.179(4)$ & $4.181(7)$ & $4.182(6)$  \\
$ -0.20$  &  $4.147(4)$ &$4.146(6)$ & $4.137(9)$ & $4.145(4)$ & $4.145(4)$ & $4.145(7)$ & $4.145(6)$  \\
$ -0.19$  &  $4.110(4)$ &$4.110(5)$ & $4.098(8)$ & $4.109(4)$ & $4.109(3)$ & $4.108(7)$ & $4.109(6)$  \\
$ -0.18$  &  $4.075(3)$ &$4.076(5)$ & $4.067(8)$ & $4.075(3)$ & $4.073(3)$ & $4.075(7)$ & $4.077(5)$  \\
$ -0.17$  &  $4.041(3)$ &$4.040(5)$ & $4.032(8)$ & $4.041(3)$ & $4.039(3)$ & $4.039(6)$ & $4.044(7)$  \\
$ -0.16$  &  $4.005(3)$ &$4.007(5)$ & $3.999(7)$ & $4.006(3)$ & $4.005(3)$ & $4.005(6)$ & $4.009(7)$  \\
$ -0.15$  &  $3.974(4)$ &$3.973(5)$ & $3.968(7)$ & $3.974(3)$ & $3.970(2)$ & $3.971(6)$ & $3.975(5)$  \\
$ -0.14$  &  $3.940(3)$ &$3.942(4)$ & $3.932(7)$ & $3.939(3)$ & $3.938(3)$ & $3.939(6)$ & $3.940(5)$  \\
$ -0.13$  &  $3.908(3)$ &$3.909(4)$ & $3.902(6)$ & $3.907(5)$ & $3.906(4)$ & $3.908(5)$ & $3.909(4)$  \\
$ -0.12$  &  $3.875(4)$ &$3.877(4)$ & $3.865(6)$ & $3.879(6)$ & $3.876(4)$ & $3.874(5)$ & $3.874(4)$  \\
$ -0.11$  &  $3.843(5)$ &$3.845(4)$ & $3.831(6)$ & $3.844(6)$ & $3.847(3)$ & $3.840(5)$ & $3.843(5)$  \\
$ -0.10$  &  $3.817(7)$ &$3.814(3)$ & $3.803(5)$ & $3.817(4)$ & $3.818(3)$ & $3.812(9)$ & $3.810(5)$  \\

  \end{tabular}
  \caption{
    The ratio of the extracted NSL to SL non-global coefficients, 
    Eq.~(\ref{eq:nsl-log-test}), for different shower
    setups, together with the sum in quadrature of statistical
    uncertainties and the AIC extrapolation systematic error.
    An additional systematic uncertainty, not included as part of
    the quoted error, is the uncertainty on the determination of $\Delta
    K_1(\bar \eta)$ and its interpolation between tabulated points. 
    To some extent that uncertainty is reflected in the spread between
    showers, which should be taken as an additional overall uncertainty.
    Systematic errors also arise from the impact of the finite size of the
    rapidity window in which emissions 
    are generated, which is $|y| < 10$.
    Reducing that to $9$ $(8)$ changes the top left result from
    $4.810\pm 0.007$ to $4.827\pm
    0.006$  ($4.837\pm 0.008$), dominated by statistical
    uncertainties.
    Taken together this suggests that the numbers above should be
    supplemented by an overall systematic uncertainty that is of the order of
    $0.5\%$.
    The results are for $2C_F =C_A=3$ and $n_f = 5$.
    The PL$^{(\text{min})}_{\beta=\frac12}$ column corresponds to an
    alternative scheme for the swaps, the ``minimal'' scheme that is
    introduced in Ref.~\cite{vanBeekveld:2026vbp}.
  }
  \label{tab:ngl-results}
\end{table}

\subsubsection{Global event shapes at NNLL accuracy}
\label{sec:nnll-shapes}
We finally consider global event shapes at NNLL accuracy. These
constitute the most comprehensive all-order tests presented here:
besides the ingredients required for the NNDL and NSL tests above,
they probe the integrated hard-collinear NLO and soft-collinear NNLO terms in the
effective coupling of Eq.~\eqref{eq:aseff} discussed in Sec.~\ref{sec:from-ds-to-nnll}.
We follow the testing strategy of Ref.~\cite{vanBeekveld:2024wws}.
For an event shape $v$, with $L=\ln v$ and
$\lambda=\as L$ held fixed, we define the NNLL quality measure
\begin{equation}
  \label{eq:nnll-shapes-test}
  \Delta_{\textsc{nnll}}(\lambda)
  =
  \lim_{\as\to0}
  \frac{\lambda^2}{\as^2}
  \left[
    \frac{\ln\Sigma_\ps(v)}{\ln\Sigma_{\textsc{nnll}}(v)}-1
  \right],
\end{equation}
where $\Sigma_\ps$ is the cumulative parton-shower distribution and
$\Sigma_{\textsc{nnll}}$ is the corresponding analytic NNLL
resummation. An NNLL-accurate shower must give
$\Delta_{\textsc{nnll}}(\lambda)=0$.
The extrapolation $\as\to 0$ is performed using the new AIC strategy
based on
$\as=$ 0.002, 0.0025, 0.0032, 0.004, 0.005, 0.0064, 0.008, 0.01,
0.0125, 0.016, 0.02.

\begin{figure}[tb!]
  \centering
\includegraphics[width=0.9\textwidth]{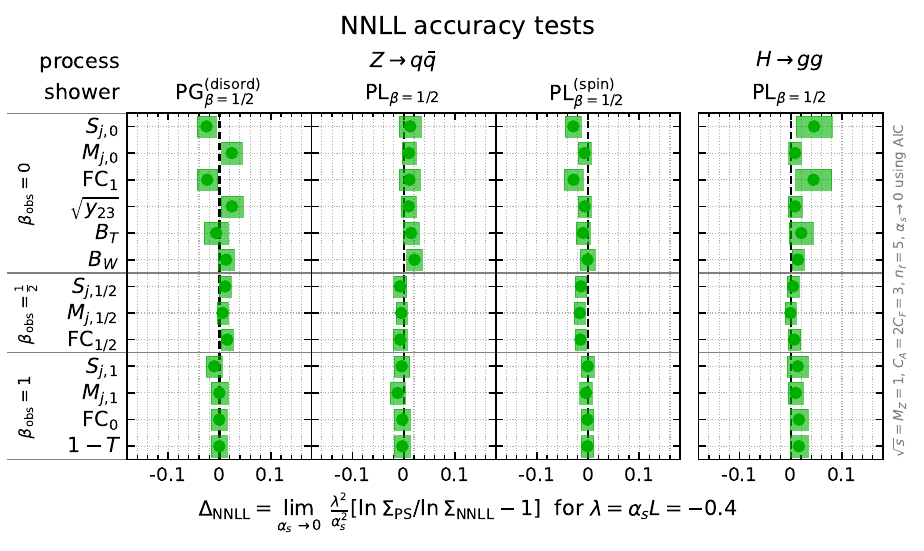}
  \caption{Summary of the NNLL tests for the global event shapes of
    Ref.~\cite{
    	Becher:2008cf,   
    	Abbate:2010xh,   
    	Chien:2010kc,    
    	Monni:2011gb,    
    	Becher:2012qc,   
    	Hoang:2014wka,   
    	Banfi:2014sua,   
    	Banfi:2016zlc,   
    	Banfi:2018mcq,   
    	Bauer:2020npd,   
    	vanBeekveld:2025zjh%
    }. The observables are grouped
    according to their soft-collinear rapidity dependence, encoded in
    $\betaobs$. The columns show the \pghalfd{} and \plhalf{} showers
    (the latter without and with spin correlations)
    for $Z\to q\bar q$, and the \plhalf{} shower for $H\to gg$
    (without spin correlations). The
    horizontal coordinate shows the size of any deviation with respect
    to the NNLL expectation (Eq.~(\ref{eq:nnll-shapes-test})), evaluated at
    $\lambda=\as L=-0.4$. All results are compatible with zero, as
    required for NNLL accuracy.}
  \label{fig:test-shapes-nnll}
\end{figure}

Fig.~\ref{fig:test-shapes-nnll} shows the result at
$\lambda=-0.4$ for several global observables. They include the Lund-tree shapes
$S_{j,\beta}$ and $M_{j,\beta}$, the $\mathrm{FC}$ observables,
the Cambridge-algorithm $\sqrt{y_{23}}$, the total and wide jet broadenings, and
thrust~\cite{
  Becher:2008cf,   
  Abbate:2010xh,   
  Chien:2010kc,    
  Monni:2011gb,    
  Becher:2012qc,   
  Hoang:2014wka,   
  Banfi:2014sua,   
  Banfi:2016zlc,   
  Banfi:2018mcq,   
  Bauer:2020npd,   
  vanBeekveld:2025zjh%
}. The observables are grouped according to
$\betaobs=0,1/2,1$, which characterises their rapidity dependence in
the soft-collinear limit.
For $Z\to q\bar q$, we test both the \pghalfd{} and \plhalf{}
showers (the latter without and with spin correlations).
For $H\to gg$, we show the corresponding \plhalf{} test (without spin correlations).
In all cases the showers include disordered emissions, together with
the complete set of NNLL corrections. All values of
$\Delta_{\textsc{nnll}}(\lambda=-0.4)$ are compatible with zero within their
uncertainties, for every observable and both processes shown.
This extends the validation of Ref.~\cite{vanBeekveld:2024wws} to the
new showers and demonstrates that the inclusion of disordered
emissions and spin correlations preserves NNLL accuracy for global event shapes.

\subsection{Timings}
\label{sec:timings}

\begin{table}
  \begin{center}
    \begin{tabular}{lccc|cccc}
      \toprule
      \multicolumn{8}{c}{ $e^+ e^- \to q\bar{q}$, $\sqrt{s} = 2$ TeV, $k_{t,\text{cut}} = 0.5$ GeV, $\alpha_s^{\MSbar}(M_Z) = 0.118$} \\
      \midrule
      with spin corr. & & & & \multicolumn{4}{c}{$\langle$event generation time$\rangle$ [$\mu$s]} \\
      \toprule
      Shower & $\betaps{}$ & disord. & $\Omega$ & NLL & 2-jet NLO & double soft & NNLL \\
      \midrule
      \pg{}           & 0   & $\times$   & 3.41  & \multirow{2}{*}{$66_{\langle N \rangle = 30}$}  & \multirow{2}{*}{$70_{\langle N \rangle = 30}$}  & $320_{\langle N \rangle = 30}$ & $380_{\langle N \rangle = 42}$ \\
      \pg{}           & 0   & \checkmark & 2.20  &                                              &                                              & $320_{\langle N \rangle = 30}$ & $390_{\langle N \rangle = 42}$ \\
      \hdashline
      \pg{}$_{\rm sdf}$ & 0   & $\times$   & 4.40 & \multirow{2}{*}{$66_{\langle N \rangle = 30}$}  & \multirow{2}{*}{$68_{\langle N \rangle = 30}$}  & $220_{\langle N \rangle = 31}$ & $460_{\langle N \rangle = 45}$ \\
      \pg{}$_{\rm sdf}$ & 0   & \checkmark & 2.20  &  &                                                                                         & $180_{\langle N \rangle = 31}$ & $380_{\langle N \rangle = 45}$ \\
      \hdashline
      \pg{}           & 1/2 & $\times$   & 22   & \multirow{2}{*}{$88_{\langle N \rangle = 30}$}  & \multirow{2}{*}{$94_{\langle N \rangle = 30}$}  & $2500_{\langle N \rangle = 35}$& $2700_{\langle N \rangle = 39}$ \\
      \pg{}           & 1/2 & \checkmark & 2.20  &                                              &                                              & $390_{\langle N \rangle = 35}$ & $440_{\langle N \rangle = 39}$ \\
      \hdashline
      \pl{}           & 1/2 & $\times$   & $-$  & \multirow{2}{*}{$100_{\langle N \rangle = 35}$} & \multirow{2}{*}{$100_{\langle N \rangle = 33}$} & $-$                         & $-$ \\
      \pl{}           & 1/2 & \checkmark & 3.00  &                                              &                                              & $600_{\langle N \rangle = 30}$ & $640_{\langle N \rangle = 32}$  \\
      \bottomrule
    \end{tabular}
  \end{center}
  \caption{The average event generation time in microseconds for all
    the PanScales showers that reach NNLL accuracy for global
    event-shapes (possibly after inclusion of disordered
    emissions). For this table we consider $e^+ e^- \to q\bar{q}$ at a
    centre-of-mass energy of $\sqrt{s} = 2$~TeV and a shower cut-off
    at $k_{t,\text{cut}} = 0.5$ GeV. On the left we report the shower,
    $\betaps{}$, whether disordered emissions are allowed, and the
    double-soft overhead $\Omega$. On the right we first show the pure
    NLL shower, then successively including 2-jet NLO matching, the
    double-soft corrections, and finally the full NNLL Sudakov. We
    also report the average particle multiplicity, $\langle N \rangle
    $, for each run. The timings were obtained compiling the code with
    the Intel$\circledR$ compilers version 2026.1.1 and \texttt{-O3}
    optimisation on an
    Intel$\circledR$~Core\texttrademark{}~Ultra~7~365 processor
    running on a single core, and by generating 100k events for each
    setup. Since timings can fluctuate even on the same machine, we
    report 2 significant digits without any further uncertainty.
    \logbook{}{%
    Timing script:
    2020-eeshower/analyses/double-soft-pheno/timings.py
    Timing results:
    2020-eeshower/analyses/double-soft-pheno/results-2026-08-alexander/timings
    Logbook :
    2026-10-09-timing-studies    
  }
}\label{tab:timings}
\end{table}
The inclusion of double-soft corrections, and higher-order kernels in
general, impacts the generation time of our showers. In
Table~\ref{tab:timings} we show all of the PanScales showers that reach
NNLL accuracy for global event shapes (with or without inclusion of disordered
emissions). The setup is always $e^+ e^- \to q\bar{q}$ at a
centre-of-mass energy of $\sqrt{s} = 2$~TeV and a shower cut-off at
$k_{t,\text{cut}} = 0.5$~GeV. We use $\alpha_s^{\MSbar}(M_Z) =
0.118$. For these timing tests we use the NODS colour-scheme with 
spin correlations turned on.
For each of the showers we show the
average event-generation time, including successively more
perturbative ingredients. The first column (NLL) is the pure NLL
shower; the second column (2-jet NLO) includes the hard matching; the
third column (double soft) includes additionally the real and virtual
double-soft correction; and the fourth column (NNLL) finally also
includes the NNLL Sudakov. As a subscript to each timing we report the
average particle multiplicity, $\langle N \rangle$ for that
shower. In general we see that the two $\betaps{}=1/2$ showers are
slower than the two variants of the \pg{}($\beta_\text{PS}=0$)
shower. The hard matching corrections impact the generation time very
little, which is perhaps not surprising as they are only ever applied for the
first emission. The double-soft corrections have a much larger
impact.
Several elements enter: a larger generation overhead, the need to
evaluate the double-soft matrix element as well as the effective
shower matrix element for multiple histories, and a larger emission
rate, which increases the multiplicity.
Finally we see
that including the full NNLL Sudakov further increases the timings
moderately, mainly because of a further increase in the emission
rate.

Allowing for disordered emissions reduces the generation
time for \pg{}($\betaps{}=1/2$) by a factor $\sim 6$. This is not surprising since the
overhead in this case is a factor $10$ smaller. The reason we do not see
a reduction in generation time by the same factor 10 is that
generating the extra disordered emissions takes additional time. We
observe that the \pl{} shower is in general slower than the \pg{}
family with double-soft corrections. This is in part due to the need
for disordered emissions, but also because the \pl{} shower has more histories
that have to be considered for the double-soft
corrections. 

One interesting observation relates to the average particle
multiplicities, and how they change across the various accuracies. In
particular we note a substantial increase in the \pg{} showers' multiplicities
when including the full NNLL Sudakov, as compared to the NLL showers.
The reason for this can be traced back to a larger value of the
effective coupling in 
Eq.~\eqref{eq:aseff}, keeping in mind that the particle multiplicity scales roughly
as $e^{\sqrt{\alpha_sL^2}}$.
At NLL the showers all use exactly the same
effective coupling, namely the CMW coupling, and as such produce
similar levels of soft radiation.
However, at NNLL the effective coupling includes contributions from
the shower-dependent drifts, which differ in the soft-wide angle
region.
As a consequence the rate of soft emissions differs between the
showers, most notably deep in the infrared.
We verified that after hadronisation with the
Pythia8~\cite{Bierlich:2022pfr} Lund string model, the $30{-}40\%$
differences in parton-level multiplicities between the
PG$_{\beta=0}^\text{(sdf)}$ and PL$_{\beta=1/2}$ NNLL showers
reduce to a $\sim 4\%$ difference in hadron-level multiplicities,
using the same \texttt{pg00sdf-nnll-2024A}
tune~\cite{vanBeekveld:2024wws} across all showers.
\logbook{}{
aa
1:# ./build-double/main-ee -shower panglobal -physical-coupling -matching BornPlus1 -split-dipole-frame -nnll-sudakov -no-spin-corr -nloops 3 -hadron-level -panscales-tune pg00sdf-nnll-2024A -nev 1e4 -rts 2000 -out aa
49:# <hadronlevel_multiplicity_unweighted> = 135.6032 +- 0.4841749245 (n entries = 10000, xsection = 0.17568386347 pb), range=[34 -- 367]

bb
1:# ./build-double/main-ee -shower panlocal -beta 0.5 -physical-coupling -matching BornPlus1 -nnll-sudakov -no-spin-corr -nloops 3 -hadron-level -panscales-tune pg00sdf-nnll-2024A -nev 1e4 -rts 2000 -out bb
49:# <hadronlevel_multiplicity_unweighted> = 129.8879 +- 0.44980376355 (n entries = 10000, xsection = 0.17596280259 pb), range=[35 -- 355]
aa-parton
1:# ./build-double/main-ee -shower panglobal -physical-coupling -matching BornPlus1 -split-dipole-frame -nnll-sudakov -no-spin-corr -nloops 3 -hadron-level no -panscales-tune pg00sdf-nnll-2024A -nev 1e4 -rts 2000 -out aa-parton
54:# <multiplicity_unweighted> = 45.143 +- 0.17142290886 (n entries = 10000, xsection = 0.1793681311 pb), range=[11 -- 131]

bb-parton
1:# ./build-double/main-ee -shower panlocal -beta 0.5 -physical-coupling -matching BornPlus1 -nnll-sudakov -no-spin-corr -nloops 3 -hadron-level no -panscales-tune pg00sdf-nnll-2024A -nev 1e4 -rts 2000 -out bb-parton
54:# <multiplicity_unweighted> = 31.9893 +- 0.11878766603 (n entries =10000, xsection = 0.1793681311 pb), range=[6 -- 92]
}%
We note that there are various interesting questions to be explored
concerning the interplay between parton shower and hadronisation
models, for example around the question of the structure of the shower's
IR cutoff in physical variables versus shower
variables~\cite{Holguin:2026trd}.

\section{Conclusions}
\label{sec:conclusion}
In this paper we have introduced a new way of organising correlated
radiation in a parton shower. Rather than requiring every
configuration to arise from a sequence that is ordered in the shower
evolution variable, we allow a second emission to be inserted above
the scale of the preceding one.
The formulation is such that insertion is restricted to an
order-one neighbourhood in logarithmic phase space and supplies only
the part of the exact double-soft rate that is not already generated
by the ordered histories.
This contrasts with earlier explorations of disordered
emission~\cite{Giele:2011cb}.
It consequently starts to affect global
event shapes at NNLL order, while leaving the NLL properties of the
underlying shower intact.

This reorganisation removes an important practical obstruction to
double-soft matrix-element corrections.  In particular, using just
strictly-ordered histories results in infinite double-soft overhead
factors for the \pl shower, and large overheads for the \pg shower
with $\betaps > 0$. 
Disordered emissions fill the corresponding phase-space regions using
modest overhead factors: in the implementations 
studied here we use $\Omega=3$ for
\pl{}($\betaps{}=1/2$) and $\Omega=2$ for
\pg{}($\betaps{}=1/2$), compared with
$\Omega\gtrsim20$ for an ordered-only implementation of the latter.

We have also revisited the treatment of collinear spin correlations. 
By using the collinear limit of the helicity-dependent branching
amplitudes to generate the branching azimuth, the shower effective
matrix element acquires a pure cosine modulation, free of the other harmonics that
could appear outside the collinear limit.
This form factorises across the histories entering the double-soft
correction and thus permits the spin and double-soft ingredients to be
combined consistently, whereas our previous work~\cite{FerrarioRavasio:2023kyg} had
implemented double-soft ingredients at the expense of
single-logarithmic accuracy for spin-sensitive observables.

The construction has been tested both at fixed order and at all
orders.  The fixed-order results reproduce the double-soft matrix
element across the relevant phase space, separately in flavour and
colour-flow channels and in the presence of spin correlations.  In
the all-order tests, the new \pg{} and \pl{} showers reproduce NNDL
Lund subjet multiplicities and NSL non-global energy flow.  After
including the shower-specific anomalous dimensions and the remaining
NNLL ingredients, they also agree with NNLL resummations for the broad
set of global event shapes and partonic channels considered here.
These tests show that disordered insertions provide the missing
correlated radiation without compromising the targeted logarithmic
accuracy.

The present study is restricted to massless final-state radiation at
leading colour.  Extensions to hadron collisions, finite-mass effects
and subleading colour would test the generality of the approach in
more complex settings.  More broadly, the separation between ordinary
ordered evolution and local disordered insertions can be applied also
outside singular regions, see e.g.\ \cite{vanBeekveld:2026vbp}.
Developing such extensions, together with fully differential
triple-collinear corrections, is a natural next step towards
general-purpose parton showers with systematically higher logarithmic
accuracy.

\section*{Acknowledgements}
We are grateful to our PanScales collaborators (Sam Bates, Mrinal
Dasgupta, Basem El-Menoufi, Jack Helliwell, Petr Jakubcik, Matvei
Kuzmin, Marco Leitão, Pier Monni,  Nicolas Schalch, Alba Soto-Ontoso,
and Silvia Zanoli) for their work on the code, the underlying
philosophy of the approach and comments on this manuscript.
GS would like to thank Laurent Lellouch for interesting discussions
about extrapolations and the AIC criterion.
This research was supported
by the Dutch Research Council (NWO) under project number
VI.Veni.232.190 (MvB),
by the Italian Ministry of Universities and Research (MUR) under the
FIS grant (CUP:D53C24005480001, FLAME) (SFR),
by the European Research Council (ERC) under the European Union’s
Horizon 2020 research and innovation programme (grant agreement
No. 788223, PanScales),
by the Australian Research Council (ARC) through a Discovery
Early Career Researcher Award (project number DE230100867) (LS),
by the French Agence Nationale de la Recherche (grant
ANR-25-CE31-7192, P2S2) (GS),
by a Royal Society Research Professorship
(RP$\backslash$R1$\backslash$231001) (GPS),
and by the Science and Technology Facilities Council (STFC)
under grants ST/X000761/1, UKRI4236 (GPS), ST/X000516/1 and UKRI4244
(KH). 
MvB, GPS and GS would like to thank the CERN Theory Department for
hospitality while part of this work was carried out. LS would like to
thank the Max Planck Institute for Physics for hospitality while this
work was being finalised.

\appendix


\section{Derivation of the collinear anomalous dimension}
\label{app:b2-nnll}

In this appendix, we revisit the derivation of the anomalous dimension $B_2$ for
the PanGlobal shower and illustrate how to derive the PanLocal one. 
We consider a hard leg $\ell$ of
energy $E$ that branches into two partons carrying energy fractions $z$ and $1-z$, with
$\theta$ the angle between the two daughter partons. 
As elsewhere in this paper, the soft limit corresponds to $z\to0$.

For a quark leg, the inclusive NLO accurate collinear splitting
probability can be written as
\begin{align}
  \label{eq:B2-Pcoll-q}
  \mathcal{P}_{{\rm coll},q}(z,\theta) =
  & \frac{\alpha_s^{\overline{\rm MS}}(zE\theta)}{2\pi}\Bigg[
    \frac{2C_F}{z}\left(1+\frac{\alpha_s^{\overline{\rm MS}}(zE\theta)}{2\pi}K_1\right) \nonumber \\
  & \qquad \qquad \quad
    + B_{1,q}(z) + \frac{\alpha_s^{\overline{\rm MS}}(zE\theta)}{2\pi}
    \Big(B_{2,q}(z) + 2\beta_0\ln z\, B_{1,q}(z)\Big)\Bigg],
\end{align}
where $K_1$ is given in Eq.~\eqref{eq:K1}, $\beta_0$ in Eq.~(\ref{eq:DeltaK2-beta0}), $B_{1,q}(z) = -C_F(2-z)$, and $B_{2,q}(z)$ is the sum of the various colour-channel contributions given in Eqs.~(3.16), (3.23), (3.21) and (3.45) of Ref.~\cite{Dasgupta:2021hbh} (there with $z\to1-z$).
The integral of $B_{2,q}(z)$ over $z$ is given in Eqs.~(24) and (25) of the supplemental material of Ref.~\cite{vanBeekveld:2024wws}.

For a gluon leg, either daughter can become soft, and the inclusive splitting probability therefore involves the coupling at the two scales~\cite{vanBeekveld:2023lsa}
$zE\theta$ and $(1-z)E\theta$,
\begin{multline}
  \label{eq:B2-Pcoll-g}
  \mathcal{P}_{{\rm coll},g}(z,\theta) = \frac{\alpha_s^{\overline{\rm MS}}(zE\theta)}{2\pi}\Bigg[
  \frac{C_A}{z}\left(1+\frac{\alpha_s^{\overline{\rm MS}}(zE\theta)}{2\pi}K_1\right)
  \\
\hspace{9.6em}  + \frac{\alpha_s^{\overline{\rm MS}}((1-z)E\theta)}{\alpha_s^{\overline{\rm MS}}(zE\theta)}
  \frac{C_A}{1-z}\left(1+\frac{\alpha_s^{\overline{\rm MS}}((1-z)E\theta)}{2\pi}K_1\right)
  \\
  + B_{1,g}(z) + \frac{\alpha_s^{\overline{\rm MS}}(zE\theta)}{2\pi}
	\Big(B_{2,g}(z) + 2\beta_0\ln z\, B_{1,g}(z)\Big)\Bigg],
\end{multline}
with
\begin{equation}
  \label{eq:B2-B1g}
  B_{1,g}(z) = C_A\big(z(1-z)-2\big) + n_fT_R\big(z^2+(1-z)^2\big).
\end{equation}
The function $B_{2,g}(z)$ can be obtained by considering Eqs.~(7.4), (7.6), (7.8), (7.10) and (7.13) (the latter without the clustering contribution) of Ref.~\cite{vanBeekveld:2023lsa} (again there with $z\to 1-z$).
The integral of $B_{2,g}(z)$ over $z$ is given in Eqs.~(32) and (33) of the supplemental material of Ref.~\cite{vanBeekveld:2024wws}.

In the PanScales showers, the collinear limit of the inclusive dipole splitting
probability can be obtained from Eqs.~\eqref{eq:emission-probability},
\eqref{eq:aseff}, giving 
\begin{equation}
  \label{eq:B2-Pps}
  \mathcal{P}_{{\rm ps},\ell}(z,\theta) = \frac{\alpha_s^{\overline{\rm MS}}(k_t^{\rm ps})}{2\pi}
  \left[1+\frac{\alpha_s^{\overline{\rm MS}}(k_t^{\rm ps})}{2\pi}
    \left(K_1 + B^{\rm ps}_{2,\ell}(z)\right)\right]P_\ell(z) + \mathcal{O}(\alpha_s^3),
\end{equation}
where $P_\ell(z)$ is the sum of the shower's LO splitting functions of
Eq.~\eqref{eq:emission-probability} for a hard leg of flavour $\ell$,
\begin{equation}
  \label{eq:B2-Pps-split}
  P_q(z) = P_{q\to qg}(z)\,, \quad
  P_g(z) = P_{g\to gg}(z) + P_{g\to q\bar q}(z)\,.
\end{equation}
In the PanScales showers, we use
\begin{equation}
  \label{eq:B2-Pps-split-explicit}
  P_{q\to qg}(z) = C_F\frac{1+(1-z)^2}{z}\,, \,\,\,
  P_{g\to gg}(z) = \frac{C_A}{2z}\left(1+(1-z)^3\right), \,\,\,
  P_{g\to q\bar q}(z) = n_fT_R(1-z)^2\,.
\end{equation}
The scale of the coupling used by the shower is denoted by $k_t^{\rm ps}\equiv k_t^{\rm ps}(z,\theta)$. 
Its expression in terms of $z$ and $\theta$ differs for PanGlobal and
PanLocal, which in the infra-red become $k_t^{\rm PG} = Ez\theta$ and $k_t^{\rm PL} = Ez(1-z)\theta$.
For a gluon leg, the splitting functions have been partitioned, breaking the
$z\leftrightarrow (1-z)$ symmetry. 
The fully symmetric collinear limit is restored after
taking into account that a gluon leg belongs to two dipoles, with the second dipole bringing
a term $\mathcal{P}_{{\rm ps},g}(1-z,\theta)$. 
The shower's effective gluon splitting
probability is therefore
\begin{equation}
  \label{eq:B2-Peff}
  \mathcal{P}_{{\rm eff},g}(z,\theta) = \mathcal{P}_{{\rm ps},g}(z,\theta) + \mathcal{P}_{{\rm ps},g}(1-z,\theta)\,.
\end{equation}
At NNLL accuracy for event shapes, only the integral of $B^{\rm ps}_{2,\ell}(z)$ enters. 
Additionally  $B^{\rm ps}_{2,\ell}(z)$ must vanish in the soft limit ($z\to 0$). 
We can therefore  define, without loss of generality
\begin{equation}
  \label{eq:B2-zB2}
  B^{\rm ps}_{2,\ell}(z) \equiv zB^{\rm ps}_{2,\ell}\,,
\end{equation}
and obtain $B^{\rm ps}_{2,\ell}$ by requiring
\begin{subequations}
  \label{eq:B2-cond}
  \begin{align}
    \label{eq:B2-cond-q}
    \int_0^1 dz \left(\mathcal{P}_{{\rm ps},q}(z,\theta) - \mathcal{P}_{{\rm coll},q}(z,\theta)\right) &= \mathcal{O}(\alpha_s^3),\\
    \label{eq:B2-cond-g}
    \int_0^1 dz \left(\mathcal{P}_{{\rm eff},g}(z,\theta) - \mathcal{P}_{{\rm coll},g}(z,\theta)\right) &= \mathcal{O}(\alpha_s^3).
  \end{align}
\end{subequations}
The coefficient $B_{2,q}$ for the quark case follows directly from Eq.~\eqref{eq:B2-cond-q}. 
For a gluon leg, as soft divergences at $z\to0$ and $z\to1$ cancel between the two terms of
Eq.~\eqref{eq:B2-cond-g}, we may expand
\begin{equation}
	\label{eq:B2-as-expansion}
	\alpha_s^{\overline{\rm MS}}((1-z)E\theta) = \alpha_s^{\overline{\rm MS}}(zE\theta)
	\left(1 - 2\beta_0\frac{\alpha_s^{\overline{\rm MS}}(zE\theta)}{2\pi}\ln\frac{1-z}{z}\right)
	+ \mathcal{O}(\alpha_s^3).
\end{equation}
Using this in Eq.~\eqref{eq:B2-cond-g}, we obtain for
PanGlobal\footnote{%
  Notice that Eq.~\eqref{eq:B2-PG-g} differs from the expression originally presented in the supplemental material of Ref.~\cite{vanBeekveld:2024wws}.
  The discrepancy arises from an inconsistent treatment of the
  $z\leftrightarrow 1-z$ symmetry in the original derivation.
  It will be discussed further in a forthcoming Erratum to
  Ref.~\cite{vanBeekveld:2024wws}.
}
\begin{subequations}
  \begin{align}
    \label{eq:B2-PG-q}
    B_{2,q}^{\rm PG} &= \frac{C_F(-27+36\pi^2-432\zeta_3) + C_A(-45+26\pi^2+216\zeta_3) + 4n_fT_R(9-4\pi^2)}{96}\,,\\
    \label{eq:B2-PG-g}
    B_{2,g}^{\rm PG} &= \frac{12C_Fn_fT_R - 2C_An_fT_R(7+2\pi^2) + 8n_f^2T_R^2 + C_A^2(-10+11\pi^2-36\zeta_3)}{15C_A + 2n_fT_R}\,.
  \end{align}
\end{subequations}
The difference between the PanGlobal and PanLocal showers is entirely driven by the choice
of renormalisation scale in the showers, specifically
$k_t^{\rm PL}(z,\theta) = (1-z)\,k_t^{\rm PG}(z,\theta)$. 
Using this we obtain for PanLocal
\begin{subequations}
  \begin{align}
    \label{eq:B2-PL-q}
    B_{2,q}^{\rm PL} &= B_{2,q}^{\rm PG} + \frac{\beta_0}{2}\left(\frac{15}{4}-\pi^2\right),\\
    \label{eq:B2-PL-g}
    B_{2,g}^{\rm PL} &= B_{2,g}^{\rm PG} + \frac23\beta_0\frac{C_A(49-12\pi^2) - 8n_fT_R}{15C_A+2n_fT_R}\,.
  \end{align}
\end{subequations}

\section{Effective parton shower matrix element}
\label{app:shower-me}
In this appendix we summarise the effective parton-shower matrix
element that enters the denominator of the double-soft correction in
Sec.~\ref{sec:ds}.
We follow the notation of Ref.~\cite{FerrarioRavasio:2023kyg}.
We consider a hard dipole $ab$ and a pair of double-soft partons,
labelled $1$ and $2$, with invariants
$s_{ij}=2p_i\cdot p_j$ and $s_i=2p_i\cdot Q$.
For a gluon pair, the two large-$\nc$ colour orderings are $a12b$ and
$a21b$; for a $q\bar q$ pair, the same notation denotes the two
assignments of the soft momenta to the quark and antiquark.
The shower contribution used in the matrix-element correction is
obtained by summing the double-soft approximation to all histories
that can give the same momenta
\begin{equation}
  \omega^{(S)} =
  \omega^{(S)}_{a12b}+\omega^{(S)}_{a21b}\, .
  \label{eq:omegaS-total-app}
\end{equation}
All expressions below are to be evaluated at fixed post-branching
momenta.  Recoil effects on the hard legs $a$ and $b$ are power
suppressed in the double-soft limit.
Note that the two terms in Eq.~(\ref{eq:omegaS-total-app}) reflect the
two possible colour assignments, which we treat simultaneously.

\subsection{PanGlobal}
\label{app:shower-me-pg}

The \pg{} expressions were given in
Ref.~\cite{FerrarioRavasio:2023kyg} and we report them again here for completeness.
The kinematic map does not distinguish between emitter and
spectator.  The effective shower matrix element with colour ordering $a21b$ can be written
as
\begin{multline}
  \omega^{(S),\mathrm{PG}}_{a21b}
  =
  (8\pi\as)^2\,2C_1\,\mathcal{J}_{a21b}^{\mathrm{PG}}
  \Big[
    f(\hat\eta_2)\,2C_2\,\delta_{g2}
    \,\,+ f(-\hat\eta_2)\,z_2 P_{\tilde 1\to12}(z_2)
    (1 + \mathcal{B}_{2g}(z_2) \mathcal{C}_{ab12})
  \Big]
  \\
  \Theta(v_{\tilde 1}>v_2)+\{1\leftrightarrow2, a\leftrightarrow b\}\, ,
  \label{eq:omegaS-PG-a21b}
\end{multline}
where the term involving the square bracket corresponds to the first row of
Fig.~\ref{fig:PG-histories}, 
and the $\{1\leftrightarrow2, a\leftrightarrow b\}$ permutation is
equivalent to the history with the same colour ordering but where $1$
is emitted after $2$, corresponding to the second row of
Fig.~\ref{fig:PG-histories}.
Here $\delta_{g2}=1$ if 2 is a gluon (i.e.\ double gluon emission), 0 otherwise (i.e.\ soft $q\bar{q}$ pair),
while $P_{\tilde 1\to12}$ corresponds to the splitting functions given
in Eqs.~(\ref{eq:B2-Pps-split-explicit}). At large $\nc$,
$C_1=C_2=C_F$ and $C_A=2C_F$.
The phase-space factor and the variables for the second branching are
\begin{align}
  \mathcal{J}_{a21b}^{\mathrm{PG}}
  &= \frac{s_{ab}}{s_{b1}(s_{a1}+s_{a2})}
     \frac{s_{a1}}{s_{a2}s_{12}}\,,
  &  \bar\eta_2
  &= \frac{1}{2}\ln\frac{s_a s_{12}}{s_1 s_{a2}}\,,
  &  z_2
  &= \frac{s_{a2}}{s_{a1}+s_{a2}}\, ,
  \label{eq:PG-J-eta-z-app}
\end{align}
where
\begin{equation}
  s_i = 2 p_i \cdot Q, \qquad s_{ij} = 2p_i \cdot p_j,
\end{equation}
with $Q$ the total four-momentum.
The ordering condition is determined from
\begin{subequations}
\begin{align}
  k_{t,\tilde 1}^{2}
  &= \frac{s_{b1}(s_{a1}+s_{a2})^2}{s_{ab}s_{a1}}\,,
  &  \rho_{\tilde 1}
  &= \left(\frac{s_a s_b}{Q^2s_{ab}}\right)^{\betaps/2}\,,
  &  \bar\eta_{\tilde 1}
  &= \frac{1}{2}\ln\frac{s_a s_{b1}}{s_b s_{a1}}\,,
  \\
  k_{t,2}^2
  &= \frac{s_{a2}s_{12}}{s_{a1}}\,,
  &  \rho_2
  &= \left(\frac{s_a s_1}{Q^2s_{a1}}\right)^{\betaps/2}\,,
  &  v_i
  &= \frac{k_{t,i}}{\rho_i}\,e^{-\betaps|\bar\eta_i|}    \,.
  \label{eq:PG-ordering-app}
\end{align}
\end{subequations}
For the standard \pg{} variants, $\hat\eta_2=\bar\eta_2$.  For the
``split-dipole frame''~(sdf) variant of Ref.~\cite{FerrarioRavasio:2023kyg},
$\hat\eta_2$ is instead the rapidity of
  parton $2$ in
the pre-branching $a1$ dipole frame, $\hat \eta_2 = \frac12
\ln\frac{s_{12}(s_{a1}+s_{a2})}{s_{a1}s_{a2}}$.
Finally, the $\mathcal{C}_{ab12}$ term encodes the spin-correlation
azimuthal modulation and is given explicitly in
Appendix~\ref{app:spin-corr-Cab}, while $\mathcal{B}_{2g}$ is either
$\mathcal{B}_{qg}$ or $\mathcal{B}_{gg}$ (Eq.~(\ref{eq:Bkg})) depending on the flavour of
parton $2$.
The $a12b$ colour flow is obtained from
Eq.~\eqref{eq:omegaS-PG-a21b} by the replacement $a\leftrightarrow b$.

\subsection{PanLocal}
\label{app:shower-me-pl}

\begin{figure}[t]
  \centering

  \includegraphics[width=0.84\textwidth]{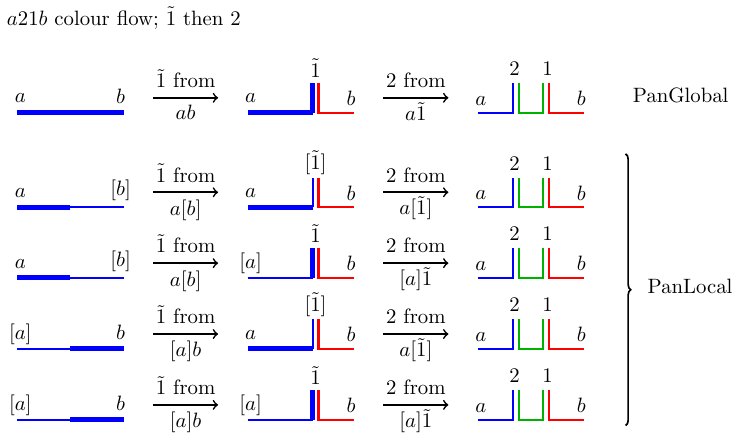}
  
  \caption{Branching histories for a double soft pair $(1,2)$ radiated
    from a dipole with hard legs $(a,b)$, in which $2$ is
    colour-connected to the hard-leg $a$ and it is emitted last. The
    first line corresponds to the case of the \pg{} shower and is
    identical to the first row of Fig.~\ref{fig:PG-histories}, the
    four bottom rows correspond to the case of the \pl{} shower where
    the four histories only differ by the choice of emitter and
    spectator in the two emissions.
    The emitter is represented by a bold line and the spectator is
    indicated in brackets.
  }
  \label{fig:PL-a21b-histories-first-four}
\end{figure}

For the \pl shower, one must additionally specify which end of a dipole is the
emitter and which is the spectator.  We write a local dipole history as
$i[j]$ when $i$ is the emitter and $j$ (in square brackets) the spectator.
For the $a21b$ colour ordering, there are eight ordered histories:
four in which parton $2$ is emitted last, illustrated in Fig.~\ref{fig:PL-a21b-histories-first-four}, and four in which $1$ is emitted last.

It is convenient to express the result as 
\begin{equation}
  \omega^{(S),\mathrm{PL}}_{a21b}
  =
  (8\pi\as)^2\,2C_1\, \sum_{r=1}^{8}
  \mathcal{J}^{\mathrm{PL}}_{r}
  \,\mathcal{K}^{\mathrm{PL}}_{r}(z_{2,r})
  \,g(\bar{\eta}_{1,r})\,g(\bar{\eta}_{2,r})
  \,\Theta(v_{1,r}>v_{2,r})\, ,
  \label{eq:omegaS-PL-a21b}
\end{equation}
where $r$ runs over the 8 possible branching histories, $v_{1,r}$ and $v_{2,r}$ are obtained from
$k_t=\rho v e^{\betaps|\bar\eta|}$ for the two branchings of
history $r$,
and the $g(\bar \eta)$ factors are the specific partitioning factors used
for the PanLocal shower (replacing the generic $f(\bar \eta)$ quoted in
Eq.~(\ref{eq:split-prob})).
The parton-shower weights associated with the four histories with
parton 2 emitted last (those of
Fig.~\ref{fig:PL-a21b-histories-first-four}) are as follows:
\begin{enumerate}
\item\label{item:first} For $a[b]\to a1b$ followed by $a[1]\to a21$
  (first \pl{} line in Fig.~\ref{fig:PL-a21b-histories-first-four}),
\begin{subequations}
\begin{align}
  \mathcal{J}^{\mathrm{PL}}_{1}
  &=
  \frac{s_{ab}}{s_{b1}(s_{a1}+s_{a2})}
  \frac{s_{a1}}{s_{a2}s_{12}}\,,
  &
  z_{2,1}&=0\,,\\
  k_{t1,1}^2
  &=
  \frac{s_{b1}(s_{a1}+s_{a2})^2}{s_{a1}s_{ab}}\,,
  &
  k_{t2,1}^2
  &=
  \frac{s_{a2}s_{12}}{s_{a1}}\,,
  \\
  \bar\eta_{1,1}
  &=
  \frac{1}{2}\ln\frac{s_a s_{b1}}{s_b s_{a1}}\,,
  &
  \bar\eta_{2,1}
  &=
  \frac{1}{2}\ln\frac{s_a s_{12}}{s_1 s_{a2}}\,,
  \\
  \rho_{1,1}
  &=
  \left(\frac{s_a s_b}{Q^2s_{ab}}\right)^{\betaps/2}\,,
  &
  \rho_{2,1}
  &=
  \left(\frac{s_a s_1}{Q^2s_{a1}}\right)^{\betaps/2}.
  \label{eq:PL-history-1-app}
\end{align}
\end{subequations}
In this case we have
\begin{equation}
  \mathcal{K}^{\mathrm{PL}}_{1}(z_{2,1})  =
  \begin{cases}
    2C_F & \mbox{ for }gg
    \\
    0 & \mbox{ for }q\bar{q}\mbox{ emissions}.
  \end{cases}
\end{equation}
\item\label{item:second} For $[a]b\to a1b$ followed by $a[1]\to a21$
  (third \pl{} line in Fig.~\ref{fig:PL-a21b-histories-first-four}),
  the expressions are as in the previous case~(\ref{item:first}) but
  with $\bar\eta_{1,2} = - \bar\eta_{1,1}$.
  Note that since this history only differs from the previous one by
  the sign of $\bar\eta_{1,r}$, one can easily compute the sum of the
  two contributions to Eq.~(\ref{eq:omegaS-PL-a21b}) using $g(\bar
  \eta_{1,1})+g(\bar\eta_{1,2})=1$.
\item\label{item:third} For $a[b]\to a1b$ followed by $[a]1\to a21$,
\begin{subequations}
\begin{align}
  \mathcal{J}^{\mathrm{PL}}_{3}
  &=
  \frac{s_{ab}}
  {(s_{b1}+s_{b2})(s_{a1}+s_{a2})-s_{ab}s_{12}}
  \frac{s_{a1}+s_{a2}}{s_{a2}s_{12}}\,,
  &
  z_{2,3}
  &=
  \frac{s_{a2}}{s_{a1}+s_{a2}}\,,
  \\
  k_{t1,3}^2
  &=
  \frac{(s_{b1}+s_{b2})(s_{a1}+s_{a2})-s_{ab}s_{12}}{s_{ab}}\,,
  &
  k_{t2,3}^2
  &=
  \frac{s_{a1}s_{a2}s_{12}}{(s_{a1}+s_{a2})^2}\,,
  \\
  \bar\eta_{1,3}
  &=
  \frac{1}{2}\ln
  \frac{s_a\left[(s_{b1}+s_{b2})(s_{a1}+s_{a2})-s_{ab}s_{12}\right]}
  {s_b(s_{a1}+s_{a2})^2}\,, \\
  \bar\eta_{2,3}
  &=
  \frac{1}{2}\ln
  \frac{s_{a2}\left[(s_{a1}+s_{a2})(s_1+s_2)-s_a s_{12}\right]}
  {s_a s_{a1}s_{12}}\,,
  \\
  \rho_{1,3}
  &=
  \left(\frac{s_a s_b}{Q^2s_{ab}}\right)^{\betaps/2}\,,
\\
  \rho_{2,3}
  &=
  \left(
  \frac{s_a\left[(s_{a1}+s_{a2})(s_1+s_2)-s_a s_{12}\right]}
  {Q^2(s_{a1}+s_{a2})^2}
  \right)^{\betaps/2}.
  \label{eq:PL-history-2-app}
\end{align}
\end{subequations}
In this case we have
\begin{equation}
  \mathcal{K}^{\mathrm{PL}}_{3}(z_{2,3})
  = z_{2,3} P_{\tilde{1}\to 12}(z_{2,3})
    [1 + \mathcal{B}_{2g}(z_{2,3}) \mathcal{C}_{ab12}].
  \end{equation}
  where $\mathcal{C}_{ab12}$ encodes the spin-correlation azimuthal
  modulation, as discussed below in Appendix~\ref{app:spin-corr-Cab}.
  
\item\label{item:fourth} For $[a]b\to a1b$ followed by $[a]1\to a21$,
  the expressions are as in the previous case~(\ref{item:third}) but
  with $\bar\eta_{1,4} = - \bar\eta_{1,3}$. Again, we can use $g(\bar
  \eta_{1,3})+g(\bar\eta_{1,4})=1$ to combine this history with the
  previous one.
\end{enumerate}
Finally, the four histories in which parton $1$ is emitted last are
obtained from the above four by the replacement
$a\leftrightarrow b$ and $1\leftrightarrow2$.

The $a12b$ colour ordering is obtained, as for \pg{}, by interchanging
$a$ and $b$.

\section{Collinear-projected spin-induced modulation factor  $\mathcal{C}_{ijkl}$}
\label{app:spin-corr-Cab}

Here we summarise the coefficients entering the double-soft effective
matrix element when (collinear-projected) spin correlations are included,
as given in Eq.~\eqref{eq:M-spin-final}.
We consider a double-soft configuration from a dipole $ab$ with
two soft emissions $k$ and $l$, with $l$ emitted last, from the $a k$ dipole.
Relative to the notation of Appendix~\ref{app:shower-me}, one should
substitute $a,b  \to i,j$ and $1,2 \to k,l$.

The momentum sharing fraction is the same for the PanGlobal and
PanLocal showers, and is
given by
\begin{equation}
	\label{eq:z2-def-spin}
	z_{2} = \frac{s_{il}}{s_{ik}+s_{il}}\,,
\end{equation}
The coefficients are shower- and history-dependent:
\begin{subequations}
\begin{align}
\text{(PanGlobal)}\qquad
\mathcal{C}_{ijkl} \,&=\, 2\left[\frac{s_{il}s_{jk}-s_{ik}s_{jl}+s_{ij}s_{kl}}
{2\sqrt{s_{ij}\,s_{il}\,s_{jk}\,s_{kl}}}\right]^{2}-1\,,
\label{eq:Cij-PanGlobal}\\[4pt]
\text{(PanLocal)}\qquad
\mathcal{C}_{ijkl} \,&=\, 2\left[\frac{\gamma_{ijkl}(-s_{ik}-s_{il})
+\beta_{ijkl}(s_{ik}-s_{il})}
{2\sqrt{\beta_{ijkl}\,s_{ij}\,s_{ik}\,s_{il}\,s_{kl}}}\right]^{2}-1\,,
\label{eq:Cij-PanLocal}\\
\beta_{ijkl} \,&=\, (s_{ik}+s_{il})(s_{jk}+s_{jl})-s_{ij}s_{kl}\,,
\nonumber\\
\gamma_{ijkl} \,&=\, s_{ik}s_{jk}-s_{il}s_{jl}\,.
\nonumber
\end{align}
\end{subequations}
They are expressed in terms of the invariants $s_{ij} = 2 p_i \cdot
p_j$.

\section{AIC strategy for the $\alpha_s\to 0$ extrapolation}
\label{app:extrapolation-strategy}

Our logarithmic tests all involve the $\as \to 0$ limit of some ratio
or difference of the shower with respect to a reference of given
accuracy. 
For example, 
the NNLL tests require the $\alpha_s \to 0$ limit of the quantity $\Delta_{\rm NNLL}(\lambda,\alpha_s)$,
with $\lambda := \alpha_s L = -0.4$ fixed,
which is evaluated at a finite set of values of the strong coupling
$\{\alpha_{s,i}\}_{i=1}^{n}$. We then fit or interpolate $\Delta_{\rm NNLL}(\lambda,\alpha_s)$ with
a polynomial $P_d$ of degree $d$ in $\alpha_s$, 
\begin{equation}
  P_d(\alpha_s)=\sum_{k=0}^{d}a^{(d)}_k\,\alpha_s^k\,,
\end{equation}
whose constant term $a^{(d)}_0$ is the quantity of interest (the intercept at $\alpha_s=0$).
Each determination $\Delta_i\equiv \Delta_{\rm NNLL}(\lambda,\alpha_{s,i})$ comes from a shower run
with Monte Carlo statistical uncertainty $\sigma_i$.
In the past we have used different methods to choose the degree $d$ of the polynomial,
as well as the values of $\alpha_s$ included in the fit. We have also used different
methods to estimate the systematic uncertainty of the extrapolation, sometimes by
varying the degree $d$ and sometimes by varying also the range of $\alpha_s$ values included
in the fit. Here we adopt a slightly more systematic approach.

In general, for finite values of $\alpha_s$, the approach to the $\alpha_s\to 0$ limit
is not guaranteed to be linear (this statement depends on the specific process, observable and
shower under consideration). Choosing too low a polynomial degree may bias the extracted value of the intercept,
because the approach may have a significant curvature still, while too high a degree fits
statistical noise and unnaturally inflates the uncertainty.
Rather than fixing $d$ by hand, here we average over
degrees with weights derived from the Akaike information criterion
(AIC)~\cite{Akaike:1974vps}, along the lines of the model-averaging procedures now standard in
lattice QCD~\cite{Jay:2020jkz}.

For each degree $d$ we perform an uncorrelated $\chi^2$ fit,
\begin{equation}
  \chi^2_d=\sum_{i=1}^{n}\frac{\big( P_d(\alpha_{s,i})-\Delta_i \big)^2}{\sigma_i^{2}}\,,
\end{equation}
and read off the extrapolated value $\Delta_d\equiv a^{(d)}_0$. The fit is a linear least-squares
problem, and so the pure statistical uncertainties $\sigma_i$ propagate to the
coefficients through
\begin{equation}
  V^{(d)}=\big(A^{\rm T}\Sigma^{-1}A\big)^{-1}\,,
  \qquad
  A_{ik}=\alpha_{s,i}^{\,k}\ \ (k=0,\dots,d)\,,
  \qquad
  \Sigma={\rm diag}(\sigma_1^2,\dots,\sigma_n^2)\,,
\end{equation}
so that $\sigma_d=\sqrt{V^{(d)}_{00}}$ is simply the statistical uncertainty of the input points
propagated to $\alpha_s=0$.

The information criterion for a fit with $k=d+1$ parameters is then
\begin{equation}
  {\rm AIC}_d=\chi^2_d+2k+\frac{2k(k+1)}{n-k-1}\,,
\end{equation}
where the first two terms are the usual AIC for a Gaussian likelihood (for which
$\chi^2=-2\ln\mathcal{L}$ up to a constant), and the last term is a small-sample
(AICc) correction, which is not negligible for the number of points used here.
Each polynomial degree is assigned a weight
\begin{equation}
  w_d=\exp\!\Big[-\tfrac12\big({\rm AIC}_d-{\rm AIC}_{\min}\big)\Big]\,,
  \qquad
  {\rm AIC}_{\min}=\min_d {\rm AIC}_d\,,
\end{equation}
normalised such that the preferred degree has $w_d=1$. The extrapolation is then quoted as
\begin{equation}
  \Delta=\frac{1}{W}\sum_d w_d\,\Delta_d\,,
  \quad
  \sigma^2_{\rm stat}=\frac{1}{W}\sum_d w_d\,\sigma_d^2\,,
  \quad
  \sigma^2_{\rm syst}=\frac{1}{W}\sum_d w_d\,\Delta_d^2-\Delta^2\,,
  \quad
  W=\sum_d w_d\,.
\end{equation}
The final statistical error is the weighted root-mean-square of the individual
fit errors,
while the systematic error is the standard deviation of the intercepts $\Delta_d$, weighted by $w_d$, across polynomial degrees.
The statistical and systematic uncertainties are added in quadrature. In practice we use
$d=1,\dots,3$.

\begin{figure}[th!]
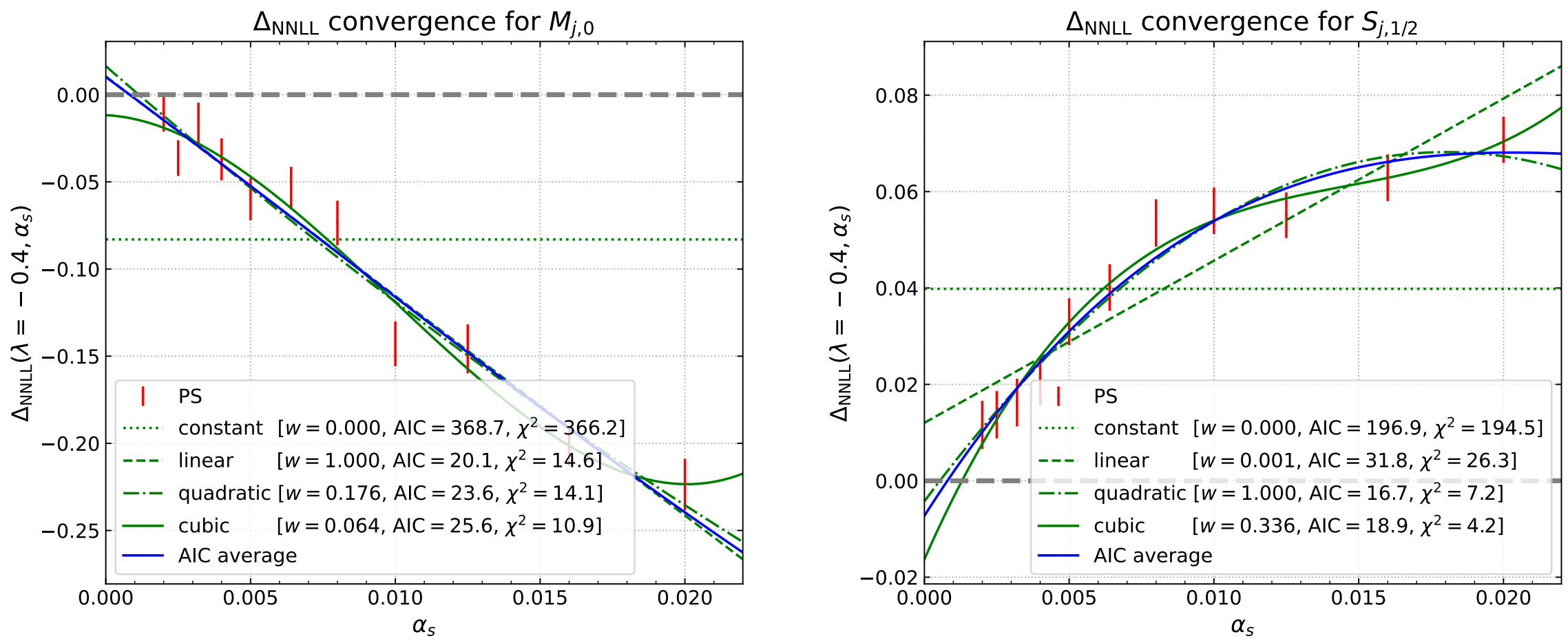

  \centering
  \includegraphics[page=2,width=0.48\textwidth]{figs/aic-illust.pdf}
  \hfill
  \includegraphics[page=7,width=0.48\textwidth]{figs/aic-illust.pdf}
  \caption{
	Illustration of the $\alpha_s\to 0$ extrapolation procedure for two observables, $M_{j,0}$ (left) and $S_{j,1/2}$ (right).
	The panels show the shower discrepancy to the NNLL result (PS, red) for different values of $\alpha_s$, with statistical error bars,
	together with the fits for degrees $d=0,\dots,3$ and the model-averaged curve (AIC, blue).
	The grey dashed line indicates the NNLL expectation, $\Delta_{\rm NNLL}=0$.
  }
  \label{fig:aic-illustration}
\end{figure}

This is illustrated in Fig.~\ref{fig:aic-illustration} for two observables, where degrees $d=0,\dots,3$
are shown (the constant is included for illustration only) with the corresponding $\chi^2$,
AIC and weight. For $M_{j,0}$ the data are compatible with a linear approach: the higher degrees
bring no significant $\chi^2$ improvement and are penalised for their extra parameters, so the
linear fit dominates. For $S_{j,1/2}$ the curvature is instead clearly visible: the linear fit is
strongly disfavoured, and the quadratic carries most of the weight,
with a non-negligible contribution from the cubic fit. In both
cases the
averaged curve (blue) lies close to the preferred fit; the alternative fits contribute to the
systematic uncertainty.

\bibliographystyle{JHEPmax9}
\bibliography{MC}

\end{document}